\documentclass[prb,showpacs,amsmath,twocolumn,amssymb]{revtex4} 

\usepackage{graphicx}
\usepackage{psfrag}
\usepackage{amsmath}
\usepackage{dcolumn}
\usepackage{bm}

\renewcommand{\Im}{\textrm{Im}\,}

\newcommand{\Jeff}{J_{\rm eff}}
\newcommand{\Hsd}{H_{\rm sd}}
\newcommand{\Hband}{H_{\rm band}}

\newcommand{\Eq}[1]{{Eq.~(\ref{#1})}}

\newcommand{\ba}{\begin{array}}
\newcommand{\ea}{\end{array}}
\newcommand{\bc}{\begin{center}}
\newcommand{\ec}{\end{center}}

\newcommand{\eps}{\epsilon}

\newcommand{\UA}{\uparrow}
\newcommand{\DA}{\downarrow}

\newcommand{\ket}[1]{{\left| #1 \right>}}

\newcommand{\biggreen}[1]{{ \big<\!\big< {#1} \big>\!\big> }}

\newcommand{\cdag}{c^{\dagger}}
\newcommand{\cnod}{c^{\phantom{\dagger}}}
\renewcommand{\ddag}{d^{\dagger}}
\newcommand{\dnod}{d^{\phantom{\dagger}}}

\begin{document}

\title{Singular Dynamics of Underscreened Magnetic Impurity Models}
\author{Winfried Koller}\email{w.koller@imperial.ac.uk}
\author{Alex C. Hewson}\email{a.hewson@imperial.ac.uk}
\author{Dietrich Meyer}\email{d.meyer@imperial.ac.uk}
\affiliation{Department of Mathematics, Imperial College, London SW7 2BZ,
  United Kingdom}

\begin{abstract}
We give a comprehensive analysis of the singular dynamics and of the
low-energy fixed point of one-channel impurity s-d models with ferromagnetic
and underscreened antiferromagnetic couplings.
We use the numerical renormalization group (NRG) to perform calculations at
$T=0$. The spectral densities of the one-electron Green's functions 
and t-matrices are found to have very sharp cusps at the Fermi
level ($\omega=0$), but do not diverge.
The approach of the Fermi level is governed by terms proportional to
$1/\ln^2(\omega/T_0)$ as $\omega\to 0$.
The scaled NRG energy levels show a slow convergence as
$1/(N+C)$ to their fixed point values, where $N$ is the iteration
number and $C$ is a constant dependent on the coupling $J$ from which
the low energy scale $T_0$ can be deduced. We calculate also the
dynamical spin susceptibility, and the elastic and inelastic
scattering cross-sections as a function of $\omega$. The inelastic
scattering goes to zero as $\omega\to 0$, as expected for a Fermi
liquid, but anomalously slowly compared to the fully screened case. We
obtain the asymptotic forms for the phase shifts for elastic
scattering of the quasiparticles in the high-spin and low-spin channels.
\end{abstract}

\maketitle

\section{Introduction}                                    \label{sec:intro}
The breakdown of Fermi liquid behaviour in the neighbourhood of a
quantum critical point (QCP) has been observed in an increasing number
of heavy fermion materials in recent years \cite{CPSR01}, but so far has not
received an adequate explanation. The QCP is the point at
which the transition temperature $T_{\rm N}$ for long range magnetic
order, usually antiferromagnetic, is such $T_{\rm N}=0$, having been
driven down via the application of pressure or alloying to a
magnetically ordered material (see for example Refs.~\onlinecite{Loe96,Ste00}).
In these circumstances the Wilson theory for critical behaviour has to be
extended to include quantum as well as thermal fluctuations of the
order parameter. Such generalizations have been carried out
\cite{Her76,Mil93,Con93} but the results do not appear to be
consistent with the experimental observations. The experimental
evidence, for instance from neutron scattering in the heavy fermion
alloy CeCu$_{5.9}$Au$_{0.1}$  \cite{SABRC98} where $\omega/T$ scaling is
found at the QCP, indicates that the critical behaviour is associated
with  very short range fluctuations in contrast with critical
phenomena driven by purely thermal fluctuations, where the shorter
range fluctuations are only important in so far as they modify the
longer range fluctuations.

One conjecture that has been put forward as a basis for a possible
theoretical explanation is that the local moments, free from the
constraints of the magnetic order, yet not fully screened as in a
conventional $S=1/2$ Kondo model, cause singular scattering leading to
a breakdown of  Fermi liquid theory \cite{CPSR01}. Such a breakdown can
occur in impurity models with a pseudogap at the Fermi
level. This type of model has been extensively studied as a model which has a
local QCP, see e.g. Refs.~\onlinecite{BPH97,GI98,IS02,GL03,FV04pre}.
For such a model to explain the observed critical behaviour, however,
an explanation would be required for the origin of the pseudogap.

The pseudogap model is not the only form of impurity model, however,
that can lead to singular scattering. It was recognized by Coleman and
P\'epin \cite{CP03} that the underscreened s-d model has free spins at
$T=0$ that can give rise to singular scattering. The s-d model is
underscreened when $2S$, where $S$ is the quantum number of the
antiferromagnetically coupled local spin, exceeds the number of
conduction electron channels. The thermodynamical properties of this
model have been known for over 20 years from exact Bethe ansatz
calculations \cite{FW81,FL82,TW83,AFL83}. The dynamical behaviour, however,
has only been studied recently, and mainly via generalizations to SU(N)
versions of the model in the large $N$ limit, where mean field methods
can be applied together with estimates of the leading Gaussian
correction terms \cite{PG97,CP03,CP04pre}. There has also been some recent
NRG work and calculations of the spinon density of states via the
Bethe ansatz \cite{MBZAC04pre}.

The underscreened s-d model might be realizable for certain quantum
dot configurations, as discussed in Ref.~\onlinecite{PG04}.
This possibility has already been considered in a recent preprint, Ref.~\onlinecite{PC04pre}.

\begin{figure}
  \includegraphics[width=0.48\textwidth]{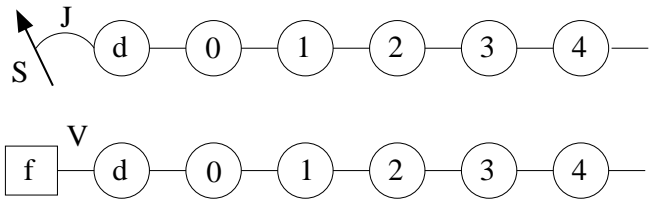}
  \caption{This figure illustrates the notation used in this paper for
  the s-d-model (top) and the Anderson model (bottom). The site
  coupling to the impurity is denoted by $d$, and in the case of the
  Anderson model the Coulomb repulsion $U$ acts at the site $f$ only.}
  \label{fig:geometry}
\end{figure}

Previous NRG calculations for the underscreened models \cite{CL79,MBZAC04pre}
have been confined to the calculation of the energy levels, using the original
Wilson procedure \cite{Wil75}. 
Here we apply the generalization of NRG approach\cite{SSK89,CHZ94} to carry
out extensive calculations of the dynamical quantities for the one-channel
underscreened model for various values of the spin $S$.
We also discuss the model with a ferromagnetically coupled spin for
comparison. The calculations are for the Hamiltonian of the s-d model
in the form, $H = \Hband + \Hsd$  with conduction electrons
\begin{equation}                               \label{eq:Hband}
  \Hband = \sum_{k,\sigma} \eps_k\,\cdag_{k,\sigma} \cnod_{k,\sigma}
\end{equation}
interacting with a localized spin $S$,
\begin{equation}
  \begin{aligned}                              \label{eq:Hint}
    \Hsd = \sum_{kk'} J_{kk'}&\big(S^+c^{\dagger}_{k,\downarrow}c^{}_{k',\uparrow}+
    S^-c^{\dagger}_{k,\uparrow}c^{}_{k',\downarrow}\\ &+
    S_z(c^{\dagger}_{k,\uparrow}c^{}_{k',\uparrow}-
    c^{\dagger}_{k,\downarrow}c^{}_{k',\downarrow})\big) \:.
  \end{aligned}
\end{equation}
We will assume a separable form for the interaction,
$J_{kk'}=J\alpha_k\alpha_{k'}$, and take
$d_{\sigma}=\sum_k \alpha_k c_{k,\sigma}^{}$
to be the site of the lattice that couples to the impurity spin,
i.e., the first site of a conduction electron chain as used in an NRG
calculation.
For $S=1/2$ and $J>0$ this is commonly known as the Kondo Hamiltonian.
We use a flat density of states $\rho_0=1/(2D)$ in the interval
$[-D,D]$ with $D=1$ and concentrate on the particle-hole symmetric case only.
The notation used in this paper is illustrated in
Fig.~\ref{fig:geometry}. It should be noted that the impurity spin is located
at site $N=-2$. 

We briefly outline the structure of the paper:
In section \ref{sec:FixedPoint} we analyse the renormalization group flows
of the low-lying excitations in terms of an effective exchange model, and
determine the renormalized coupling in terms of an effective energy
scale $T_0$.  We then calculate the spectral densities of the local
one-particle Green's functions and t-matrices for $T=0$ as a function
of $\omega$ in section \ref{sec:Greent}, the elastic and inelastic
scattering cross-sections in section \ref{sec:scattering}, and the dynamical
spin susceptibilities in section \ref{sec:spinsus}.
In the final section  we review the overall picture that these results give
into the physics of this class of impurity models, and discuss the relation
between our conclusions  and previous work on this topic.

\section{Fixed Point Analysis}                          \label{sec:FixedPoint}

As a preliminary to the calculation of the dynamics, we look at the NRG flows
and the approach to the fixed point. This will clarify the nature of the fixed
point and the form of the leading correction terms. For this analysis we use a
relatively large value of the NRG discretisation parameter $\Lambda=3.5$ and
retain $600$ states. Due to the slow convergence to the fixed point, we find
it necessary to do up to $250$ iterations.
\subsection{Ferromagnetic coupling ($J<0$)}
We begin the fixed-point analysis with the ferromagnetic case.
Here\cite{CL79} the fixed-point Hamiltonian corresponds to free fermions on
the uncoupled chain ($J=0$) in the limit $N\to\infty$ with the
additional degeneracy due to the uncoupled spin $S$. The Hamiltonian for the
uncoupled chain for finite $N$ can be diagonalized and written in the form
\begin{equation}                                         \label{eq:qp_FM}
  \Lambda^{-(N-1)/2}
  \sum_{r=1}^{(N+2)/2}\big(
  E^{(0)}_{p,r}(N)p^{\dagger}_{r,\sigma}p^{}_{r,\sigma} +
  E^{(0)}_{h,r}(N)h^{\dagger}_{r,\sigma} h^{}_{r,\sigma}\big)\:,
\end{equation}
where $p^{\dagger}_{r,\sigma}$,
$p^{}_{r,\sigma}$, and $h^{\dagger}_{r,\sigma}$, $h^{}_{r,\sigma}$,
are the creation and annihilation operators for the single particle and
hole excitations, and $\Lambda^{-(N-1)/2}E^{(0)}_{p,r}(N)$ and
$\Lambda^{-(N-1)/2}E^{(0)}_{h,r}(N)$ are the corresponding excitation
energies relative to the ground or vacuum state $|0\rangle$; the scale
factor $\Lambda^{-(N-1)/2}$ is due to the fact that the energies are
calculated for the rescaled Hamiltonian. We have assumed particle-hole
symmetry, $E^{(0)}_{p,r}(N)=E^{(0)}_{h,r}(N)$, so we can drop the 
p and h labels in denoting the levels. The levels are ordered such that
$E^{(0)}_{r+1}(N)\ge E^{(0)}_{r}(N)$ and the scaling is such that
for $r\ll N$ $E^{(0)}_{r}(N)$ is of order 1 and has a finite value
$E^*_{r}$ in the limit $N\to \infty$.

The leading correction terms to the fixed point are expected to be the
simplest local terms consistent with the symmetry of the model (see Wilson's
original paper~\cite{Wil75} for a more detailed discussion). In this
case they are expected to be of the same form as the original
Hamiltonian\cite{CL79}, and for the $N+2$ site system can be expressed in the
form,
\begin{equation}                                    \label{eq:H_c}
  \begin{aligned}
    H_c(N) =  \tilde J(N)&\big(S^+d^{\dagger}_{\downarrow}d^{}_{\uparrow}+
    S^-d^{\dagger}_{\uparrow}d^{}_{\downarrow}+
    S_z(d^{\dagger}_{\uparrow}d^{}_{\uparrow}-
    d^{\dagger}_{\downarrow}d^{}_{\downarrow})\big).
  \end{aligned}
\end{equation}
An on-site potential scattering term is excluded due to particle-hole
symmetry. 
As $H_c(N)$ describes the excitations relative to the ground state of the
uncoupled chain ($N+2$ sites), it has to be normal ordered and
written in terms of the single particle and hole operators, using 
\begin{equation}
  d^{\dagger}_{\sigma}=\sum_{r}\alpha_{0r}(p^{\dagger}_{r,\sigma}+h^{}_{r,\sigma})
\end{equation}
where the coefficients $\alpha_{0r}$ will depend upon $N$.

For the particle excitations only, $H_c(N)$ takes the form
\begin{align}
    H_c(N) =& \tilde J(N)\sum_{r,r'}
    \alpha_{0r}\alpha_{0r'}\big(S^+p^{\dagger}_{r,\downarrow}p^{}_{r',\uparrow}+
    S^-p^{\dagger}_{r,\uparrow}p^{}_{r',\downarrow}      \nonumber\\ 
    &+S_z(p^{\dagger}_{r,\uparrow}p^{}_{r',\uparrow}-
    p^{\dagger}_{r,\downarrow}p^{}_{r',\downarrow})\big)\:.
\end{align}
There is an additional similar term for the hole excitations, and another one
containing combinations of particle and hole excitations.
We will concentrate on the effects of these terms on the lowest
single-particle and two-particle excitations of the $N+2$ site system.

In the ferromagnetic case, we consider {\em even} iterations in which case
the ground state has total charge $Q=0$ and total spin $S_{\rm tot}=S$.
The lowest single particle excitations can have quantum numbers
$S_{\rm tot}=S\pm 1/2$ and $Q = 1$. The corresponding energies are denoted by
$E_{p,1}(N,S\pm 1/2)$.
As these terms become asymptotically small in the limit $N\to\infty$ we 
calculate the shift in this excitation to first order in $\tilde J(N)$.
The total energy of the scaled excitation with spin $S+1/2$ is then given by 
\begin{equation}                                   \label{eq:shift1_FM}
  E_{p,1}(N,S+1/2)=E^{(0)}_{p,1}(N)+\tilde J(N)S\alpha_{01}^2\Lambda^{(N-1)/2}
\end{equation}
and that for  spin $S-1/2$
\begin{equation}                                    \label{eq:shift2_FM}
    E_{p,1}(N,S-1/2)=E^{(0)}_{p,1}(N)-\tilde J(N)(S+1) \alpha_{01}^2\Lambda^{(N-1)/2}\:.
\end{equation}
From the fact that the high-spin excitation is found to lie lowest
we infer that the effective interaction $\tilde J(N)$ must be
ferromagnetic ($\tilde J(N) < 0$).
The energy difference between high-spin and low-spin excitations is given by
\begin{equation}                                     \label{eq:diff_FM}
  \begin{aligned}
    E_{p,1}(N,S+1/2)&-E_{p,1}(N,S-1/2)\\
    &=\tilde J(N)(2S+1)\alpha_{01}^2\Lambda^{(N-1)/2}\:.
  \end{aligned}
\end{equation}
This difference approaches its fixed point value, which is zero, very 
slowly, whereas the factor $\alpha_{01}^2\Lambda^{(N-1)/2}$ rapidly reaches
a finite fixed point value denoted by
$\bar\alpha_{01}^2 \equiv \lim_{N\to\infty} \alpha_{01}^2\Lambda^{(N-1)/2} \approx 0.27$ for
$\Lambda=3.5$.
This suggests that the asymptotic form of $\tilde J(N)$ for large $N$
is given by 
\begin{equation}                              \label{eq:Jtilde_FM}
  \tilde J(N) = - \frac{A}{N+C(J)} \:,
\end{equation}
where $A>0$ is a constant.
The slow fall off of $\tilde J(N)$ with $N$, $\tilde J(N)\propto 1/N$,
indicates that the exchange term  is a {\em marginally irrelevant}
interaction\cite{CL79}.

We can translate the $N$-dependence of the effective parameter
$\tilde J (N)$ into a frequency or temperature scale using
$\omega_N=\eta D\Lambda^{-(N-1)/2}$ or $T_N=\omega_N$, where $\eta$ is
an appropriately chosen constant of order unity\cite{Wil75,SSK89} which we take as
$\eta=1$. The relation is
\begin{equation}                                      \label{eq:N_of_omega}
  N = -\frac{2}{\ln \Lambda}\,\ln\Big(\frac{\omega_N}{\eta\,D}\Big)\:.
\end{equation}
Upon inserting this into \Eq{eq:Jtilde_FM} and expressing $C(J)$ in
terms of an energy scale $T_0$, we obtain
\begin{equation}                              \label{eq:Jtilde_FM_omega}
  \tilde J(\omega) =
  -\frac{\tilde A}{-\ln(\omega/\eta D) +  \ln(T_0/\eta D)}\:.
\end{equation}
with $\tilde A = A\,\ln \Lambda/2$ and
\begin{equation}                                  \label{eq:T0_FM}
  T_0 = \eta D \,e^{-C(J)/2\, \ln \Lambda} \:.
\end{equation}
Thus we find $\tilde J(\omega)\propto 1/\ln(\omega/T_0)$ or
$\tilde J(T)\propto 1/\ln(T/T_0)$.
In Sec.~\ref{sec:spinsus} we show that this result is in line with 
the fitting to the low frequency part of the spectrum of the dynamic
susceptibility for $T=0$. It is also in line with the exact low
temperature thermodynamics known from the Bethe
ansatz\cite{TW83,FW81,AFL83}.

For small values of $J$, the dominant contribution to $C(J)$ is
proportional to $1/J$. Hence we rewrite it in the form
\begin{equation}                              \label{eq:Jtildeb_FM}
  C(J) = - \frac{B(J)}{J} \:,
\end{equation}
with the parameter $B(J)$ only weakly dependent on $J$.
The energy scale $T_0$ reads
\begin{equation}                                  \label{eq:T0_FM1}
  T_0 = \eta D \,e^{B(J)/(2 J)\, \ln \Lambda} \:.
\end{equation}
\begin{figure}
\includegraphics[width=0.47\textwidth]{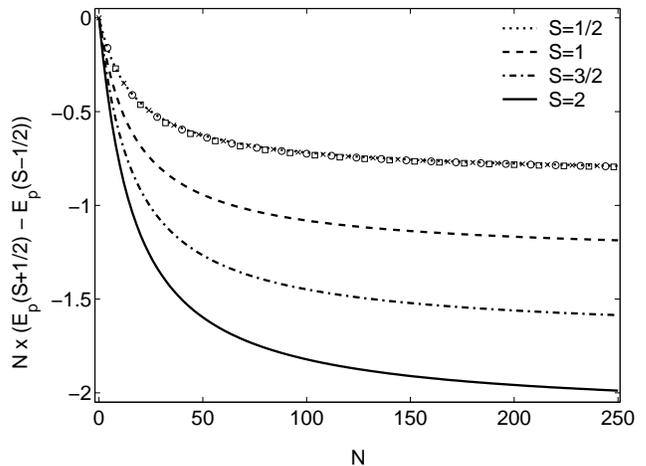}
\caption{The flow of the energy difference $N\times(E_{p,1}(S+1/2) -
  E_{p,1}(S-1/2))$ of high-spin and low-spin one-particle excitations for
  different values  of $S$ and ferromagnetic coupling $J=-0.1$.
  The symbols ($\times$ for $S=1$, $\circ$ for $S=3/2$, and $\scriptsize\square$ for
  $S=2$) show the flows rescaled by $1/(S+1/2)$.}
\label{fig:flow_1p_hl_Sxp_J20}
\end{figure}

Figure~\ref{fig:flow_1p_hl_Sxp_J20} shows the flows of the energy difference
$N\times(E_p(S+1/2) - E_p(S-1/2))$ for $J=-0.1$ and different values of $S$.
The rescaled curves (symbols) show that $A$ and $B(J)$ are indeed
independent of $S$.
We can obtain the numerical values for $A$ and $B(J)$ by inserting
\Eq{eq:Jtildeb_FM} into \Eq{eq:Jtilde_FM} and the result into
\Eq{eq:diff_FM}. Fitting this to the energy difference obtained from
the NRG, we obtain a value of $A \approx 1.56 \pm 0.01$ independent of
$J$ and $S$ for a small $J$.
The values for $B(J)$ are given in Table~\ref{tab:Jtilde_FM}.

Our value of $A \approx 1.56$ yields an $\tilde A=0.977 \approx 1$.
\Eq{eq:Jtilde_FM_omega} thus simplifies to 
\begin{equation}                              \label{eq:Jtilde_FM_omega1}
  \tilde J(\omega) =  1/\ln(\omega/T_0)\:,
\end{equation}
which agrees with the scaling result used for the spin susceptibility in
\Eq{eq:scale_J_chi}.

\begin{table}
\begin{center}
\begin{tabular}{|c|c|c|c|c|}
\hline \textbf{J} & \textbf{S=1/2} & \textbf{S=1}& \textbf{S=3/2} & \textbf{S=2}\\ 
\hline -0.005 & 1.5990 & 1.5991 & 1.5992 & 1.5994\\
\hline -0.010 & 1.6034 & 1.6036 & 1.6039 & 1.6043\\
\hline -0.025 & 1.6226 & 1.6223 & 1.6219 & 1.6214\\
\hline -0.050 & 1.6637 & 1.6586 & 1.6516 & 1.6426\\
\hline -0.100 & 1.7589 & 1.7270 & 1.6827 & 1.6261\\
\hline -0.200 & 1.9540 & 1.7970 & 1.5789 & 1.3017\\
\hline
\end{tabular} 
\end{center}
\caption{Values for $B(J)$ obtained by numerical fitting of
  the low-energy behaviour (iterations $50-250$) to \Eq{eq:Jtilde_FM} and
  \Eq{eq:Jtildeb_FM} using   $250$ iterations at $\Lambda=3.5$.} 
\label{tab:Jtilde_FM}
\end{table}

In the limit $N\to\infty$, the energy $E_{pp,1}(N,S)$ of the lowest two-particle
excitation with quantum numbers $Q=2$ and $S_{\rm  tot} = S$ is equal to
$2\times E^{(0)}_{p,1}(N)$ of the non-interacting case.
The difference $E_{pp,1}(N,S)-2E^{(0)}_{p,1}(N)$ falls off relatively slowly
with $N$, approximately as $1/N^2$, which we can interpret as a second order
effect of $\tilde J(N)$.

Following the analysis used by Hofstetter and Zar\'and \cite{HZ04} we can 
deduce the asymptotic form  the phase shifts in the two spin channels,
$\eta_{S+1/2}(\omega)$ and $\eta_{S-1/2}(\omega)$, from the energy
levels of the low-lying single particle excitations in the approach to the
fixed point ($N\to\infty$).
The relation of the phase shift to the NRG flows is
\begin{equation}                                    \label{eq:def_ps}
  E_{p,1}(N) =  E^*_{1}\,\Big(1 - \tfrac{1}{\pi}\,\eta(N)\Big)
\end{equation}
with $E^*_{1}$ denoting the one-particle excitation energy at the
fixed point.
Translating the $N$ dependence in a $\omega$ dependence and making use
of the fact that $E^{(0)}_{p,1}(N) \to E^*_{1}$ very rapidly,
we can use \Eq{eq:shift1_FM} and \Eq{eq:shift2_FM} to calculate the
phase shift.
For the $S+1/2$ channel we obtain
\begin{equation}                                   \label{eq:phaseshift1_FM}
  \eta_{S+1/2}(\omega)=-{\bar \alpha_{01}^2\over E^*_{1}}\pi S\tilde J(\omega)
\end{equation}
and similarly for the $S-1/2$
\begin{equation}                                    \label{eq:phaseshift2_FM}
   \eta_{S-1/2}(\omega)=
   {\bar \alpha_{01}^2\over E^*_{1}} \pi (S+1) \,\tilde J(\omega)\:.
\end{equation}
The numerical value of the proportionality factor is 
$\bar\alpha_{01}^2 / E^*_{1} = 0.49$ for a $\Lambda=1.8$.
In appendix \ref{app:exscatt} we consider explicitly the exchange scattering
of a single quasiparticle by an interaction $\tilde J$, which should
asymptotically describe the behaviour near the fixed point.
The resulting phase shifts are given in \Eq{eq:nps} and can be seen to give
the same as those estimated from the level shifts for $\omega \to 0$
($\tilde J(\omega) \to 0$).

\subsection{Antiferromagnetic coupling ($J> 0$)}
\begin{figure}
  \includegraphics[width=0.47\textwidth]{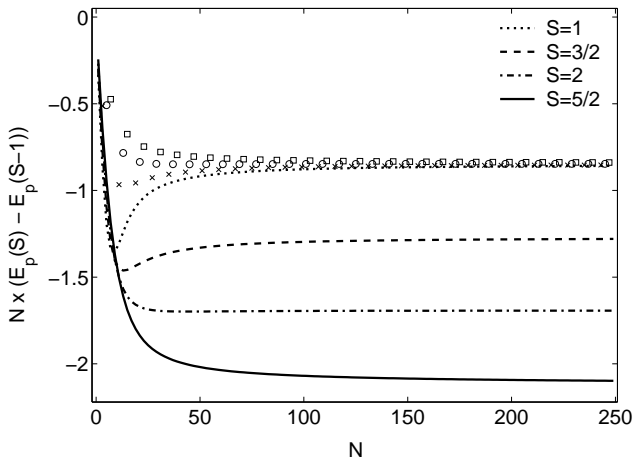}
  \caption{The flow of the energy difference $N\times(E_{p,1}(S) -
    E_{p,1}(S-1))$ of high-spin and low-spin one-particle excitations  
   for different values of $S$ and antiferromagnetic coupling
    $J=+0.2$. The symbols ($\times$ for $S=3/2$, $\circ$ for $S=2$, and
    $\scriptsize\square$ for  $S=5/2$) show the flows rescaled by
    $1/S$.}  
  \label{fig:flow_1p_hl_Sxm_J40}
\end{figure}

For antiferromagnetic coupling, a partial screening of the spin $S$ takes
place yielding a ground state with a reduced spin $S-1/2$.
This screening introduces a phase shift of $\pi/2$, and 
therefore, in order to facilitate a comparison with the ferromagnetic case,
we consider only {\em odd} iterations here.

We assume that the low-lying excitations from the ground state can
again be expressed using \Eq{eq:H_c} and calculate $\tilde J(N)$ using
the flows of the NRG.
Figure~\ref{fig:flow_1p_hl_Sxm_J40} shows the flow of
\begin{equation}                                  \label{eq:high-low_AFM}
  E_{p,1}(N,S)-E_{p,1}(N,S-1) =
  \tilde J(N)(2S)\alpha_{01}^2\Lambda^{(N-1)/2}
\end{equation}
multiplied by $N$.
This corresponds to the quantity analyzed in the ferromagnetic case as the
antiferromagnetic model has a ground state with $S$ replaced with $S-1/2$.

Looking first at the curves for smaller $S$, we see two distinct regions.
Initially, $\tilde J(N) \times N$ decreases linearly with $N$. For larger
values of $N$ we see the flows increasing relatively rapidly.
For higher values of $S$, the flow is monotonically decreasing as in
the ferromagnetic case.
This suggests that the low-energy fixed point is the same as in the 
ferromagnetic case but with an additional phase shift of $\pi/2$ and a spin
$S-1/2$ reduced by $1/2$. Also the approach to the fixed point is formally
identical, see \Eq{eq:Jtilde_FM} with $S \rightarrow S-1/2$.

\begin{figure}
  \includegraphics[width=0.47\textwidth]{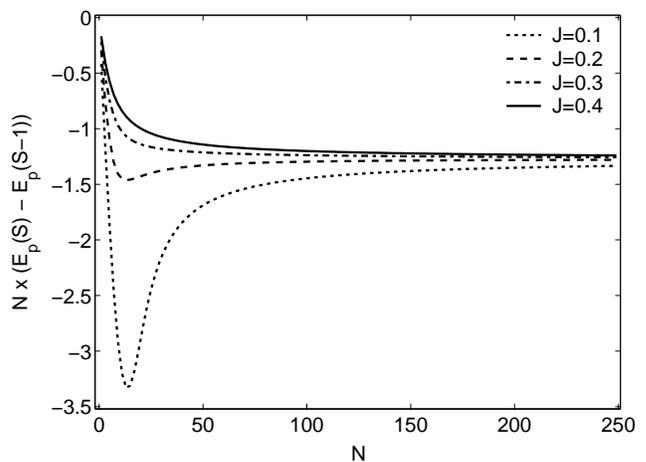}
  \caption{The flow of the energy difference $N\times(E_{p,1}(S) -
    E_{p,1}(S-1))$ of high-spin and low-spin one-particle excitations  
    for $S=3/2$ different values of the antiferromagnetic coupling $J$.}  
  \label{fig:flow_1p_hl_S3m_Jx}
\end{figure}

As can be seen in Fig.~\ref{fig:flow_1p_hl_Sxm_J40} and
\ref{fig:flow_1p_hl_S3m_Jx}, the approach to the fixed 
point can vary depending on $J$ and $S$. For $T_0/D>1$ ($T_0/D<1$),
the fixed point is approached from below (above).
Looking at Fig.~\ref{fig:flow_1p_hl_S3m_Jx}, we see that for a fixed
$S=3/2$ there is a crossover from a non-monotonic flow for smaller values
of $J$ to a monotonically decreasing flow for higher values.
In the latter case, the flows are completely similar to those with a 
ferromagnetic bare $J<0$. Therefore it makes sense to use
\Eq{eq:Jtildeb_FM} to define a effective bare $\Jeff$ for the
antiferromagnetic case.
We will see below that $\Jeff$ is usually ferromagnetic.
Then, using \Eq{eq:Jtilde_FM}, the coupling reads
\begin{equation}                                   \label{eq:Jtilde_AFM}
  \tilde J(N) = - \frac{A}{N-B(J)/\Jeff(J)}\:.
\end{equation}
The parameter $\Jeff$ can be interpreted as an effective coupling
between the residual spin and the conduction electrons, provided we are
in the strong coupling regime.
From Table~\ref{tab:Jtilde_FM} we infer that in this regime
a reasonable approximation to chose a constant $B$.
Therefore we use a fixed value $B=1.62$ in order to fit
\Eq{eq:Jtilde_AFM} to the NRG flows (\Eq{eq:high-low_AFM}).

\begin{figure}
  \includegraphics[width=0.48\textwidth]{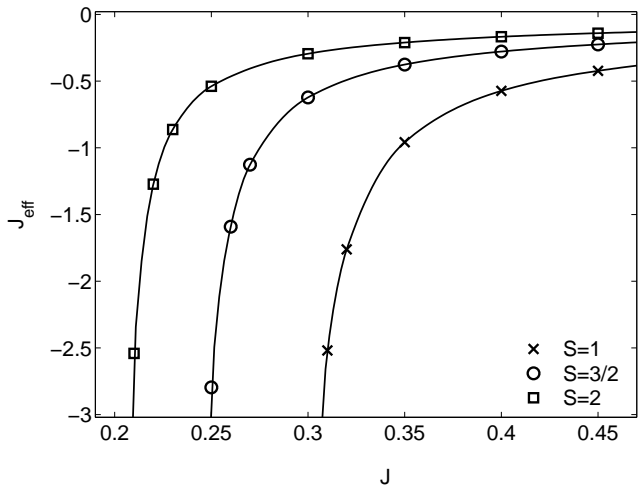}
  \caption{The effective ferromagnetic coupling $J_{\rm eff}$ as a function of the
  bare antiferromagnetic $J$ for $S=1 \,(\times)$, $S=3/2 \,(\circ)$, and $S=2
  \,(\scriptsize\square)$.} 
  \label{fig:Jeff_AFM_Sx}
\end{figure}

The fits yield the same value of $A \approx 1.57$ as in the
ferromagnetic case for all $S$ and $J$.
The results for $\Jeff$ as a function of $J$ are plotted in
Fig.~\ref{fig:Jeff_AFM_Sx}.
We observe that $\Jeff \to 0$ as $J \to \infty$ and that
$\Jeff$ depends strongly on $S$.
For smaller values of $J$ than plotted in Fig.~\ref{fig:Jeff_AFM_Sx},
$\Jeff$ is positive and hence the coupling is antiferromagnetic.
It should be pointed out, however, that
$\tilde J(N)$ is always ferromagnetic. This can also be seen in
Fig.~\ref{fig:flow_1p_hl_S3m_Jx} where all energy differences are negative.

As in the ferromagnetic case, we can derive the phase shifts in the two spin
channels. These evaluate to
\begin{equation}                                   \label{eq:phaseshift1_AFM}
  \eta_{S}(\omega)=\frac \pi 2 -
      {\bar \alpha_{01}^2 \over E^*_{1}}\pi \big(S-\frac 1 2\big)\, \tilde J(\omega)
\end{equation}
in the high-spin case and similarly 
\begin{equation}                                    \label{eq:phaseshift2_AFM}
   \eta_{S-1}(\omega)=\frac \pi 2 +
       {\bar \alpha_{01}^2 \over E^*_{1}} \pi \big(S+\frac 1 2\big)\, \tilde J(\omega)
\end{equation}
in the low-spin case. The former phase shift agrees with that calculated from
the spinon density of states in Ref.~\onlinecite{MBZAC04pre}.

\section{Green's functions and t matrices}                    \label{sec:Greent}
For the calculation of dynamic quantities, we apply the NRG with a
discretisation parameter $\Lambda=1.8$ and retain up to $1600$ states.
We focus, first of all, on the one-electron Green's functions and t-matrices.
To calculate the t-matrix for the s-d model, we take the equation of
motion for the single-electron Green's function
$ G_{k\uparrow,k'\uparrow}(\omega)=\biggreen{\cnod_{k,\uparrow}
  : \cdag_{k',\uparrow}}$
and obtain the relation
\begin{equation}                           \label{eq:G_kk}
  G_{k\uparrow,k'\uparrow}(\omega)=
  {\delta_{kk'} \over \omega -\epsilon_k} + 
  {J\alpha_{k} \over \omega-\epsilon_k}
  \,G_{t}(\omega)\,
  {J\alpha_{k'} \over \omega -\epsilon_{k'}}\:,
\end{equation}
where $G_{t}(\omega)$ is given by
\begin{equation}                                      \label{eq:def_Gt}
  G_{t}(\omega) =
  \langle\langle S^- \dnod_{\downarrow}+ S_z \dnod_{\uparrow} :
  S^+ \ddag_{\downarrow} + S_z \ddag_{\uparrow} \rangle\rangle\:.
\end{equation}
Hence in this case the {\em on-shell} t-matrix\cite{Lan66} can be expressed in
terms of the Green's function $G_{t}(\omega)$ via 
\begin{equation}
  t_{k,k'}(\omega) = J^2\,\alpha_k\alpha_{k'}\,G_{t}(\omega)\;.
\end{equation}
We can find a relation between $G_{t}(\omega)$ and the local electron Green's
function $G_{d}(\omega) \equiv \biggreen{\dnod_\sigma : \ddag_\sigma}$ on the
$d$ site.
For this we multiply \Eq{eq:G_kk} by $\alpha_k$ $\alpha_{k'}$ and
sum over all $k$ and $k'$.
We find
\begin{equation}
  G_{d}(\omega) =
  G_d^{(0)}(\omega)+ J^2\,
  G_d^{(0)}(\omega)\, G_{t}(\omega)\,  G_d^{(0)}(\omega)\:,
\end{equation}
with the non-interacting $d$-site Green's function
\begin{equation}                                       \label{eq:G_d0}
  G_d^{(0)}(\omega)=\sum_k{|\alpha_k|^2\over{\omega-\epsilon_k}}\:.
\end{equation}
In the wide band limit we can take $G_d^{(0)}(\omega)=-i\pi\rho_0$ for
$\omega \ll D$, and in this case we find
\begin{equation}
  G_{d}(\omega) = -i \pi \rho_0 - \pi^2\rho_0^2 J^2
  G_{t}(\omega)\:.
\end{equation}
From this we deduce for the spectral densities 
\begin{equation}                                         \label{eq:rho_d0t}
  \rho_{d}(\omega) = \rho_0 \big[1-\pi^2\rho_0 J^2\rho_{t}(\omega) \big]\:,
\end{equation}
where $\rho_{t}(\omega)=-\Im G_{t}(\omega)/\pi$.

We know from the results for the Anderson model (see
appendix~\ref{app:Anderson}) that $\rho_d(0)=0$.
This must apply to the s-d model with antiferromagnetic coupling with
spin $S=1/2$, because it is equivalent to the Kondo limit of the Anderson
model.
By inserting $\rho_d(0)=0$ into \Eq{eq:rho_d0t} we find
\begin{equation}                    \label{eq:rho_of_zero}
  \rho_{t}(0)={1\over \pi^2\rho_0 J^2}\:.
\end{equation}
We conjecture that this result applies quite generally for antiferromagnetic  
coupling and  all values of $S$.
This is confirmed in our numerical results given below.

\subsection{Anderson model and Kondo model}

%
\begin{figure} 
  \includegraphics[width=0.45\textwidth]{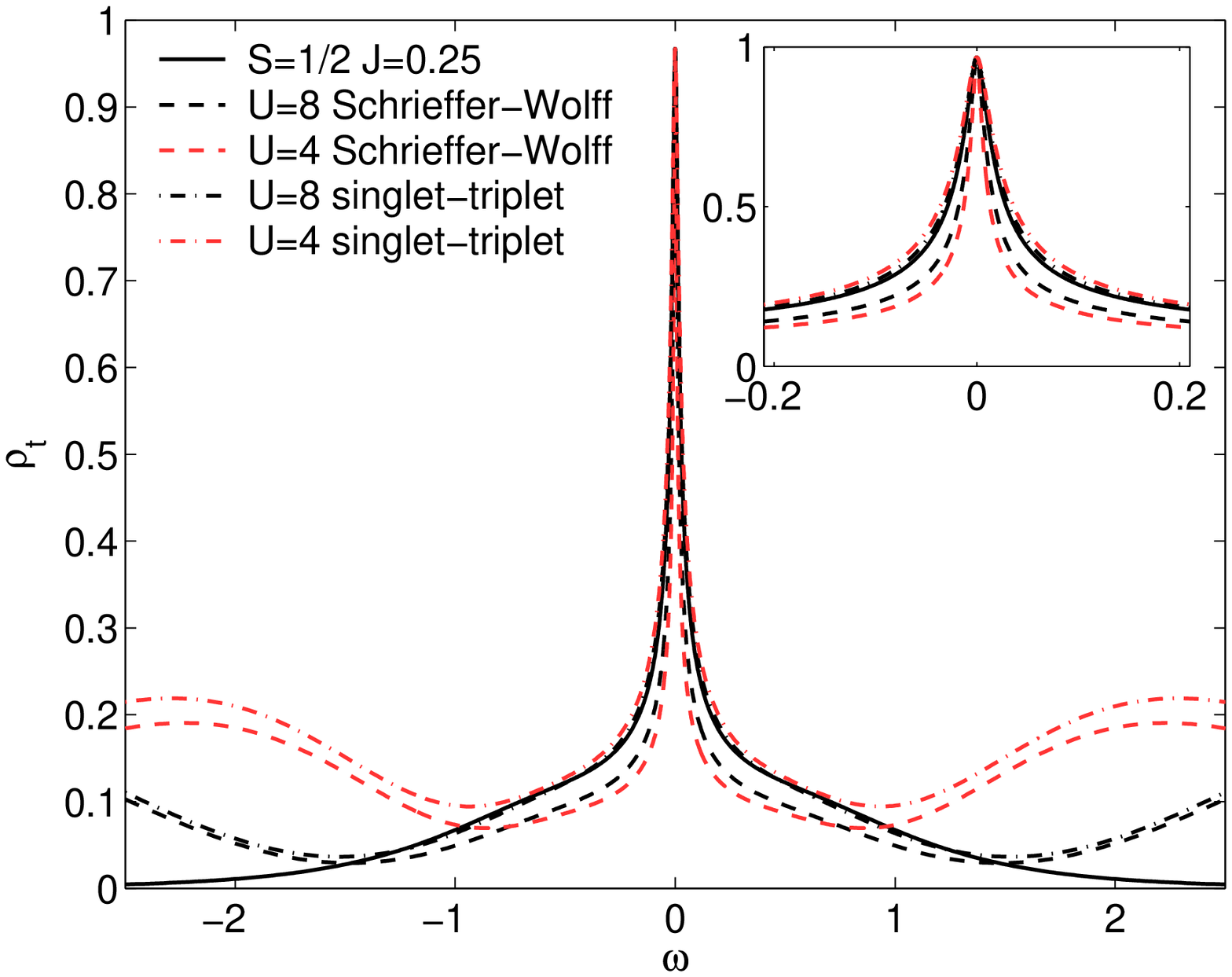} \\[1ex]
  \includegraphics[width=0.45\textwidth]{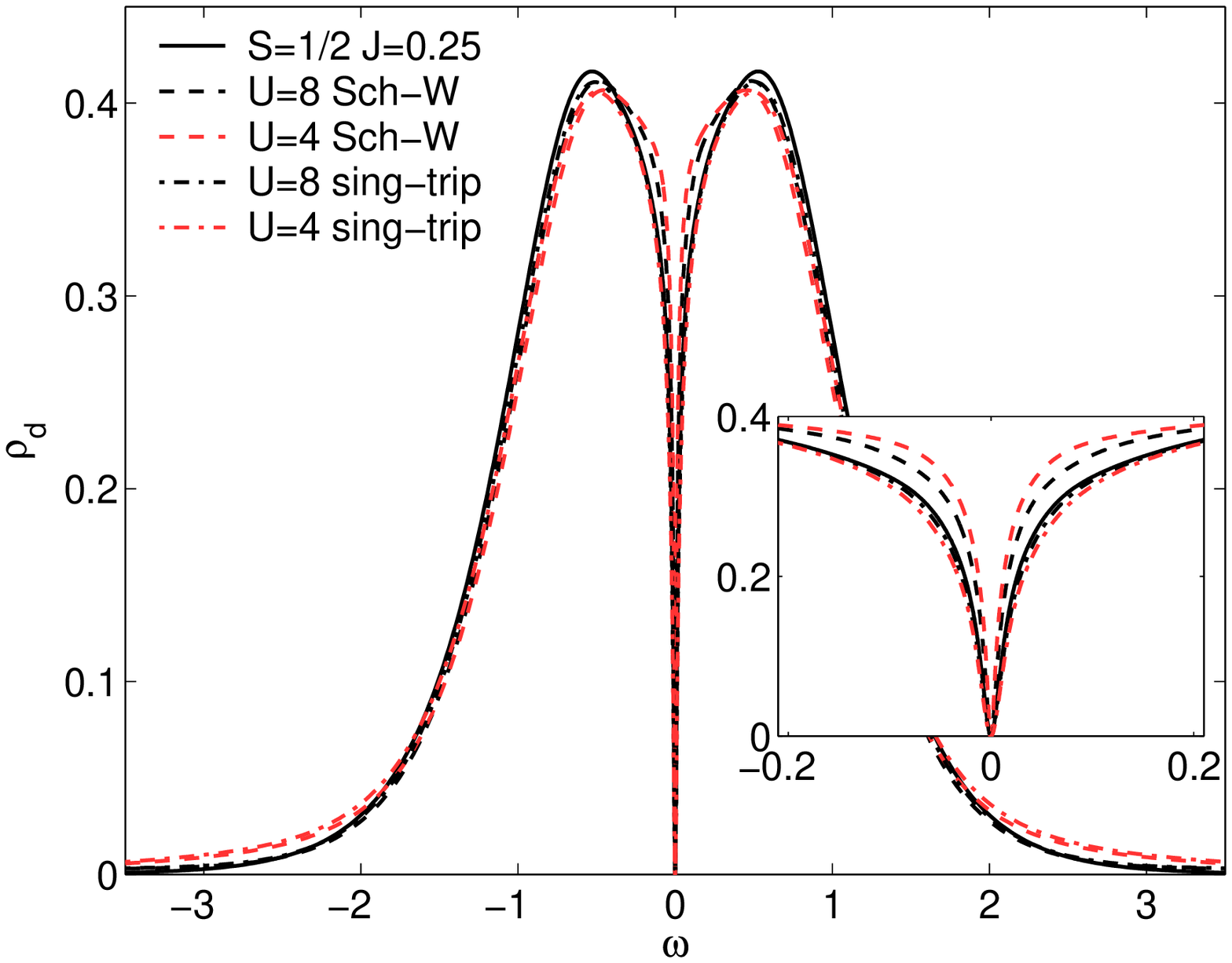}
  \caption{Spectral densities $\rho_{f}$ of the Anderson model and $\rho_{t}$
    of the Kondo model (top) as well as $\rho_{d}$ (bottom) for both model;
    A resonance in $\rho_f$ and $\rho_t$ corresponds to an antiresonance in~
    $\rho_d$.} 
  \label{fig:KM_AM}
\end{figure}

In this subsection we look at the relation between the Kondo model
(in our notation: s-d model with $S=1/2$ and antiferromagnetic coupling) and
the Anderson impurity model.
In the latter model the spin $S=1/2$ is replaced with an $f$-level
that couples to the conduction band at the site $d$ with
hybridization~$V$, see Fig.~\ref{fig:geometry}.
On this $f$-site the electrons interact with a
Hubbard repulsion $U$. The Anderson model and the Kondo model are known to be
related by the Schrieffer-Wolff transformation\cite{SW66} which yields
$J = 4 V^2 / U$. 

Another way of relating these two models is to look at the
singlet-triplet splitting at the isolated impurity in both cases.
For symmetric parameters of the Anderson model this yields
\begin{equation}                                \label{eq:sing_trip}
  J = - \frac{U}{8} + \frac{1}{2}\,\sqrt{\left(\frac{U}{4}\right)^2 +\, 4\,V^2}\:.
\end{equation}
By expanding the square root to first order in $V^2/U^2$ one recovers
the Schrieffer-Wolff result.

Figure~\ref{fig:KM_AM} compares the spectra of the two models. The
top panel shows $\rho_t$ of the Kondo model along with the spectrum
$\rho_f$ of the local $f$-site Green's function for two different
values of $U$. Dashed lines refer to a value of $V$ as calculated from
the Schrieffer-Wolff transformation, whereas dot-dashed lines are for
a $V$ as calculated from Eq.~(\ref{eq:sing_trip}). One observes that
in all cases the central resonance of the Kondo model is well
approximated by $\rho_f$ of the Anderson model. $V$ from the
singlet-triplet splitting, however, yields a slightly better
approximation to the Kondo model. 
Obviously, $\rho_{t}$ of the Kondo model does not show the charge excitations
present in the Anderson model's $\rho_f$.

The same can be said for the anti-resonance at the $d$-site shown in
the bottom panel of Fig.~\ref{fig:KM_AM}.
This anti-resonance is expected from Eq.~(\ref{eq:rho_d0t}).
Again the singlet-triplet splitting yields a somewhat better approximation to
the anti-resonance of the Kondo model than the Schrieffer-Wolff result.

\subsection{Ferromagnetic Coupling}
\begin{figure}
  \includegraphics[width=0.45\textwidth]{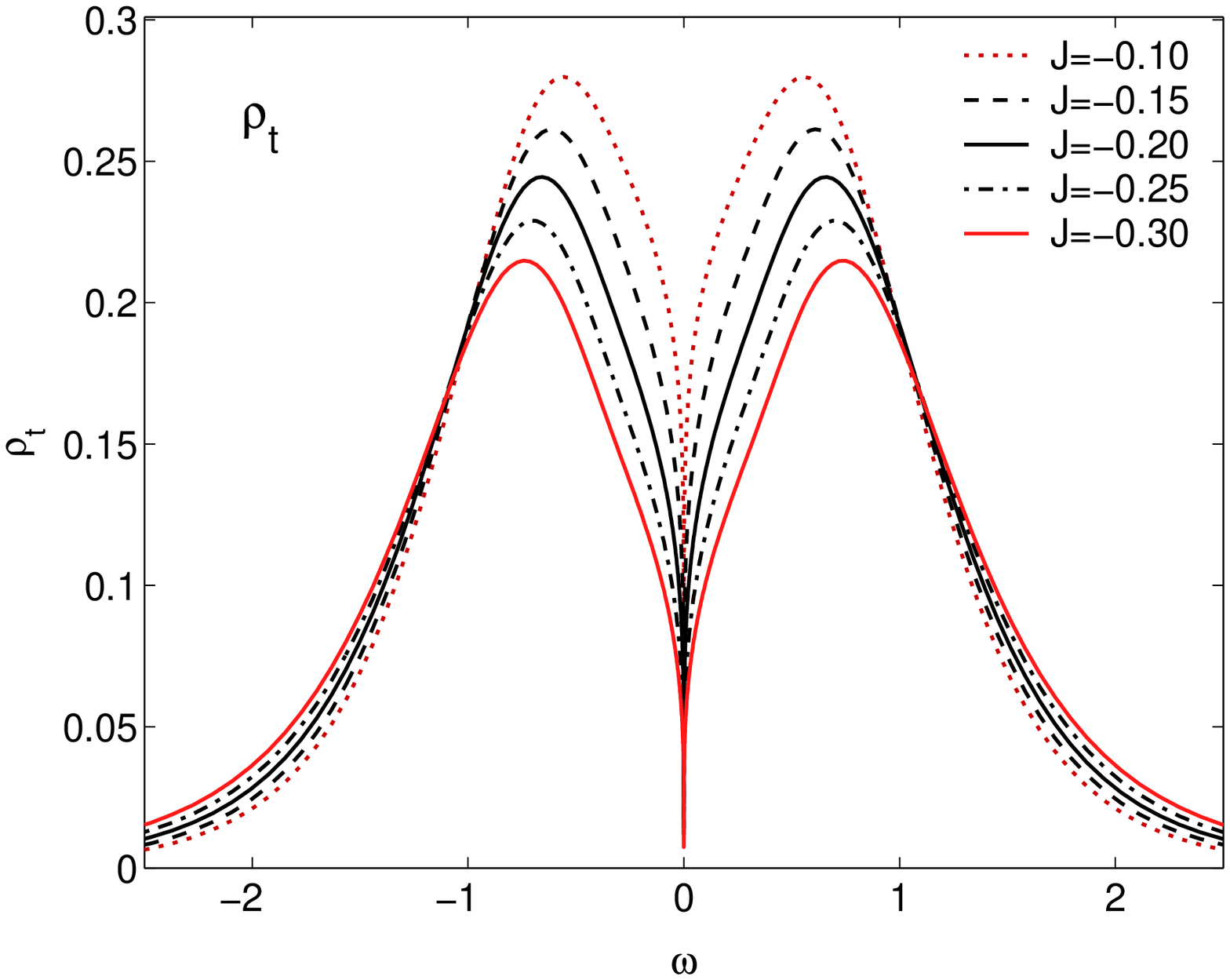}\\[1ex]
  \includegraphics[width=0.45\textwidth]{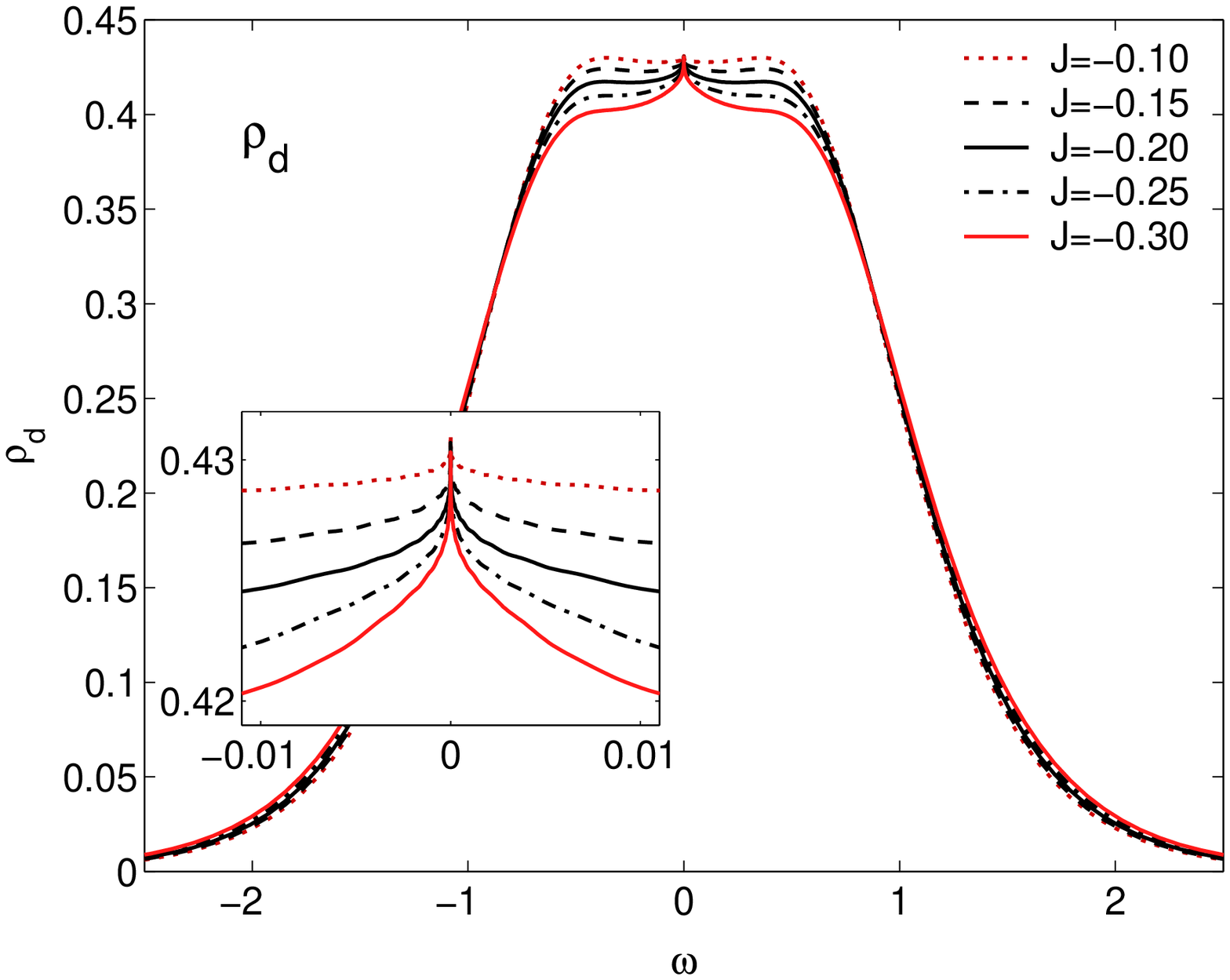}
  \caption{Spectra $\rho_{t}$ (top) and $\rho_{d}$ (bottom) for $S=1/2$ and
    various values of $J$ for ferromagnetic coupling. The
    anti-resonance in $\rho_d$ corresponds to the resonance in $\rho_t$.}
  \label{fig:gt_spec_S1p}
\end{figure}
\begin{figure}
  \includegraphics[width=0.48\textwidth]{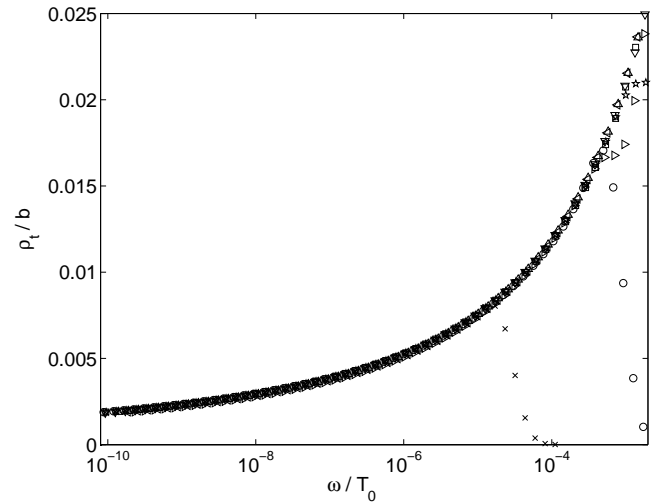}
  \caption{Low-energy behaviour of the rescaled spectral function $\rho_{t }$
  as a function of the reduced energy $\omega/T_0$ for FM coupling to a spin
  $S=1/2$.
  After rescaling, the  spectra corresponding to all coupling strengths fall
  on one curve that is well fitted by $\rho_t(\omega) = b /
  \ln^2(T_0/\omega)$. The range of fitting is always 
  the interval $[10^{-8} , 0.5]$ independent of $J$ which takes the values
  $-0.1\,(\times), -0.15\,(\circ)$ to $-0.45\,(<)$.}
  \label{fig:t_spec_rescale_S1p}
\end{figure}
\begin{figure}
  \includegraphics[width=0.48\textwidth]{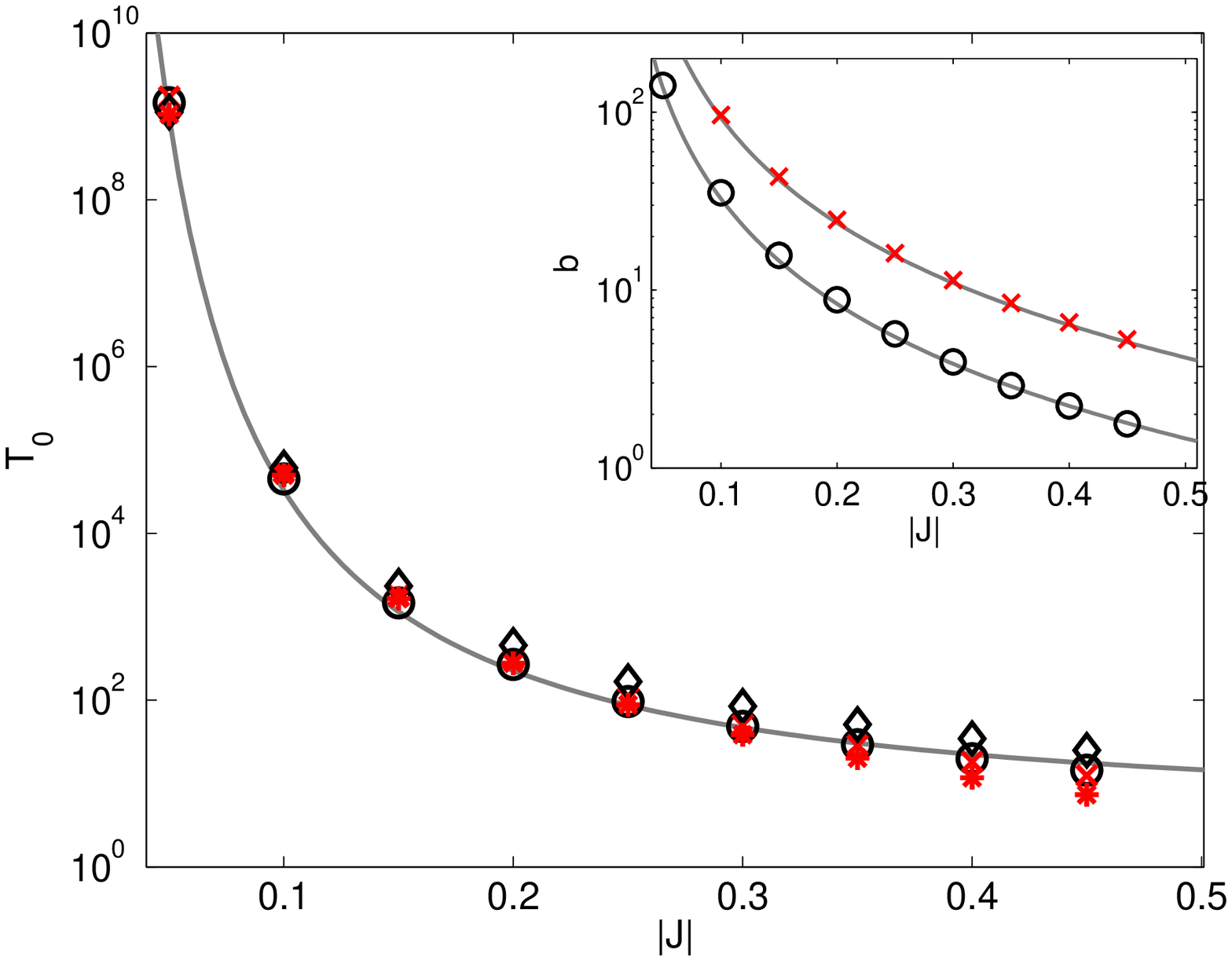}
  \caption{Energy scale $T_0$ as a function of the ferromagnetic coupling
    strength $J$ for different $S$.
    Estimates of $T_0$ from the fit of $\rho_t(\omega) = b /
    \ln^2(\omega/T_0)$ over the interval $[10^{-8}, 10^{-2}]$ are shown as 
    $(\circ)$ for $S=1/2$ and as $(\times)$ for $S=1$.
    Estimates of $T_0$ from the flows through \Eq{eq:T0_FM} are denoted by
    $\diamond$ for $S=1/2$  and $\ast$ for $S=1$.
    The full line joining the points is given by $T_0(J) =
    D\,\sqrt{2\rho_0J}\,\exp(-1/(2\rho_0J))$ with $D=2.55$ and $\rho_0=0.47$.
    In the inset the parameter $b$ is plotted as a function of $J$.
    Here the data for $S=1$ are roughly $3\, \times$ those for
    $S=1/2$. }
  \label{fig:T0_fit_Sxp}
\end{figure}

The case of a ferromagnetic coupling to a spin $S=1/2$ is completely
different from the antiferromagnetic one, since there is no tendency
to screen the impurity spin. There is no corresponding Anderson model.

Qualitatively, the shape of the spectra is reversed, as can be seen in
Fig.~\ref{fig:gt_spec_S1p}.
For ferromagnetic coupling, $\rho_{t}$ has an anti-resonance whereas
$\rho_{d}$ a resonance whose height is independent\cite{SMN02} of $J$.
This independence stems from the fact that the anti-resonance goes
straight down to $\rho_{t}(0)=0$, in which case \Eq{eq:rho_d0t} implies
$\rho_d(0) = \rho_0$.
The low-energy behaviour of $\rho_{t}(\omega)$ can be
fitted to the function
\begin{equation}                          \label{eq:rho_t_fit}
  \rho_t(\omega) \sim  \frac b {\ln^2(\omega/T_0)}
\end{equation}
which vanishes at $\omega=0$. 
This point is illustrated in Fig.~\ref{fig:t_spec_rescale_S1p} where
the rescaled spectra for various coupling strengths are plotted on a
logarithmic low-energy scale.
Equation~(\ref{eq:rho_t_fit}) also reflects the singular approach of
$\rho_t(\omega)$ to $\omega=0$ as the derivative diverges at this
point.  This singular behaviour is also seen in the cusp-shaped
peak of $\rho_d$ at $\omega=0$ (see inset to Fig.~\ref{fig:gt_spec_S1p}). 

As is shown in Fig.~\ref{fig:T0_fit_Sxp}, $T_0$ seems to be
independent of $S$ and can be fitted with the formula
\begin{equation}                           \label{eq:T0_def}
  T_0(J) = D\,\sqrt{2\rho_0J}\,e^{-1/(2\rho_0J)}
\end{equation}
with $D=2.55$ and  $\rho_0=0.47$.
This energy scale agrees remarkably well with the energy scale~$T_0$ obtained
from the fixed point analysis, see \Eq{eq:T0_FM}. In the ferromagnetic
case, $T_0$ is a large energy scale and  greater than the bandwidth $D$.

\subsection{Antiferromagnetic Coupling}

\begin{figure}
  \includegraphics[width=0.45\textwidth]{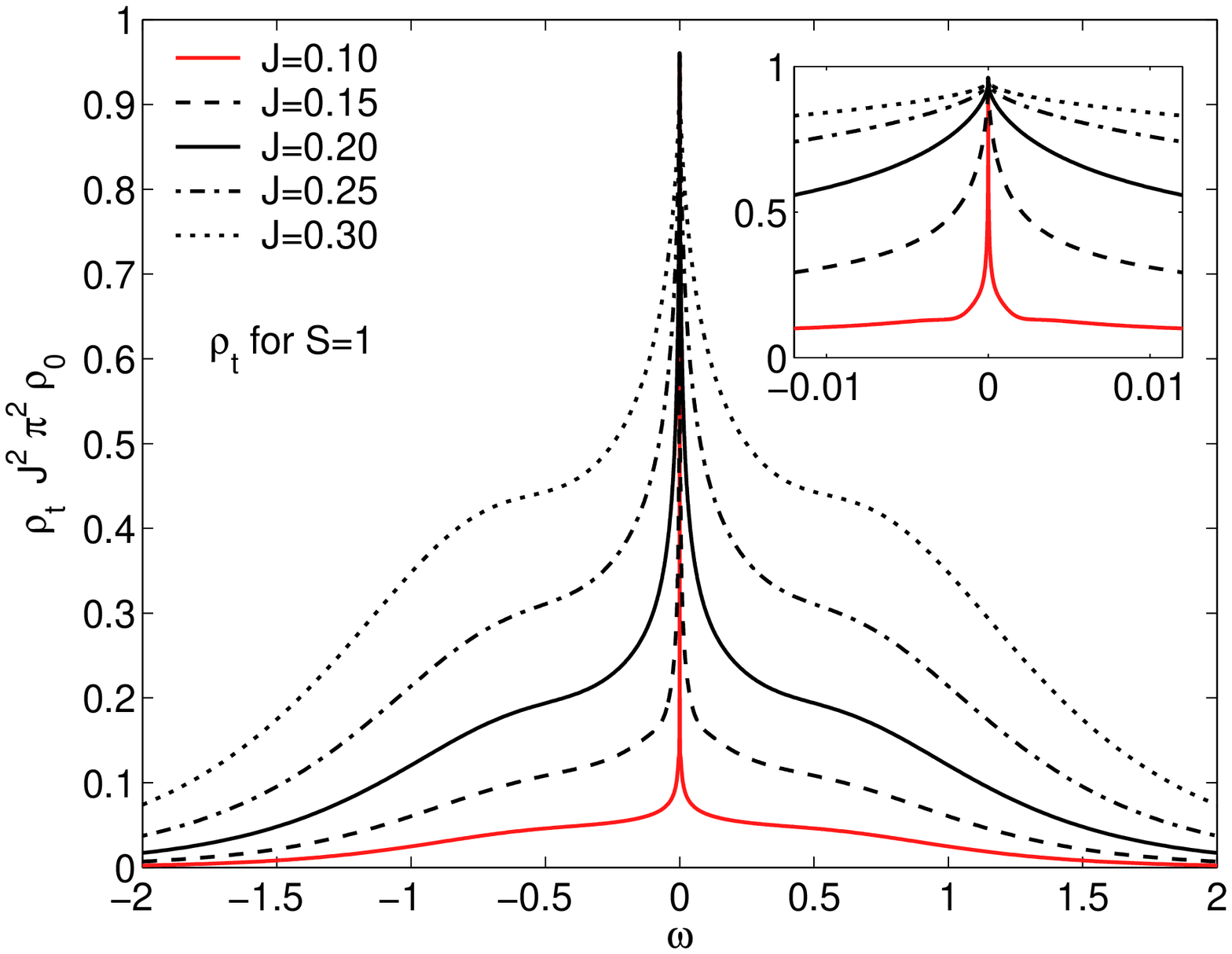}\\[1ex]
  \includegraphics[width=0.45\textwidth]{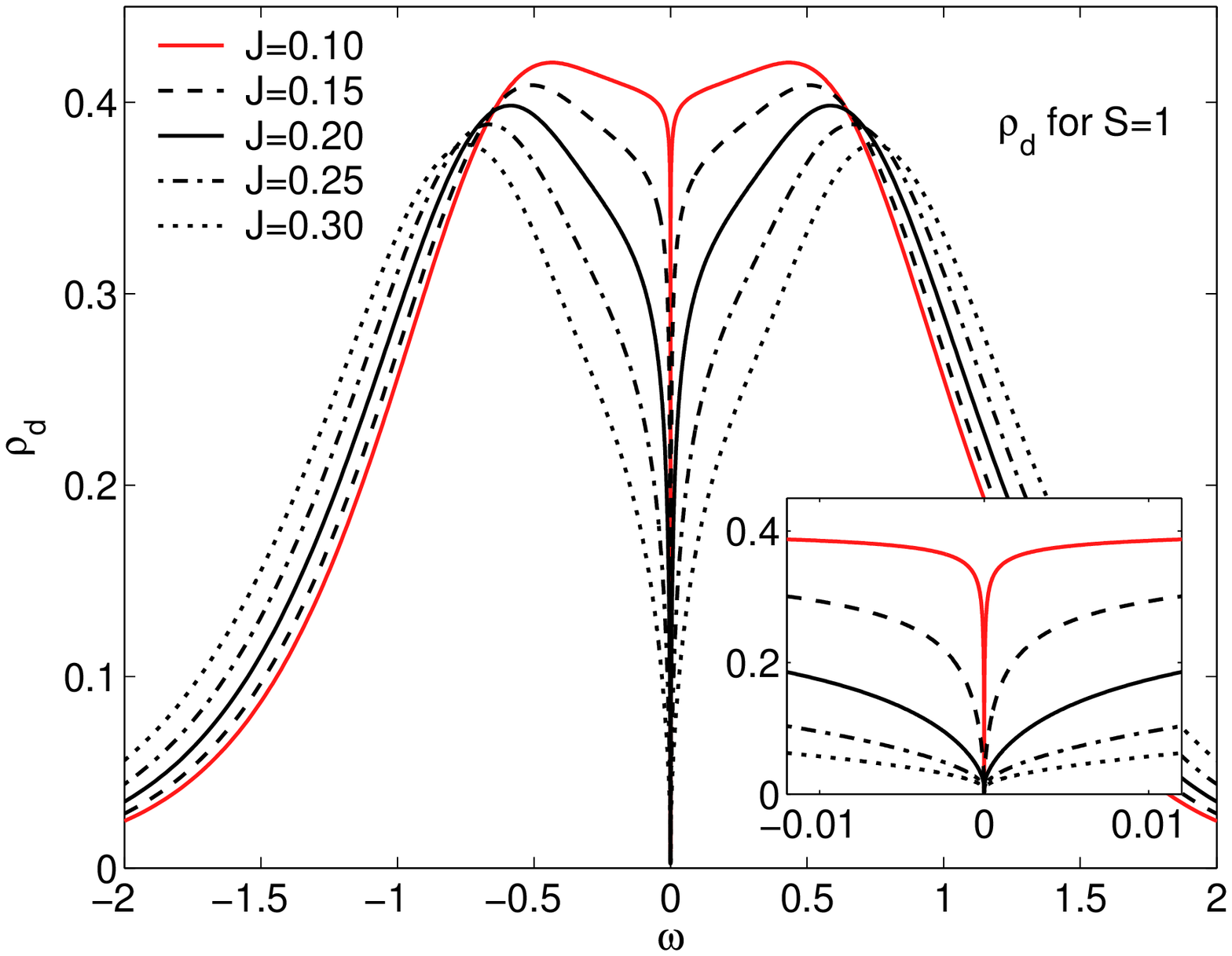}
  \caption{Spectral densities $\rho_{t}$ (top) and $\rho_{d}$ (bottom)
    of the underscreened s-d model with $S=1$ resonance and
    antiresonance; former normalized by $J^2 \pi^2 \rho_0$}  
  \label{fig:rho_G_S2_Jx}
\end{figure}

We turn to the spectra of the underscreened s-d models with
antiferromagnetic coupling to a spin $S>1/2$.
Here, a resonance at $\omega=0$ is found in $G_t(\omega)$ and an
antiresonance in $G_d(\omega)$. These are plotted in
Fig.~\ref{fig:rho_G_S2_Jx} for $S=1$ and various values of $J$.
The features at $|\omega| \approx 1$ are due to the band edge and
$\rho_t(0) \approx 1/(J^2 \pi^2 \rho_0)$ as one would expect from
\Eq{eq:rho_of_zero}.
The gap in $\rho_d(\omega)$ at $\omega=0$ remains for all $J<0$, and
$\rho_d(0)=0$ however small $J$ in contrast to the situation $J=0$, where
there is no gap and $\rho_d(\omega) = \rho_0$.

\begin{figure}
  \includegraphics[width=0.45\textwidth]{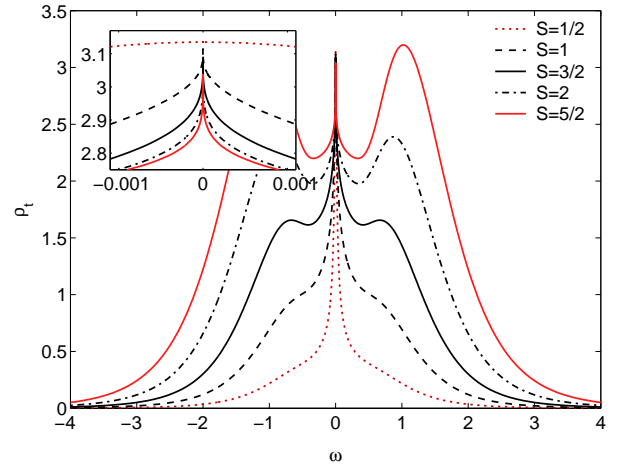}
  \caption{Spectral densities $\rho_{t}$ of the antiferromagnetically coupled
    s-d model with $S=1/2$ (dotted), $S=1$ (dashed), $S=3/2$,  $S=2$
    (dot-dashed), and $S=5/2$. 
    The inset shows the resonance on a lower energy scale, where the spectrum
    of the fully screened model (dotted) is almost flat. The value at
    $\omega=0$ is approximately the same for all values of $S$.}
  \label{fig:t_spec_Sx_Jm50}
\end{figure}

\begin{figure}
  \includegraphics[width=0.45\textwidth]{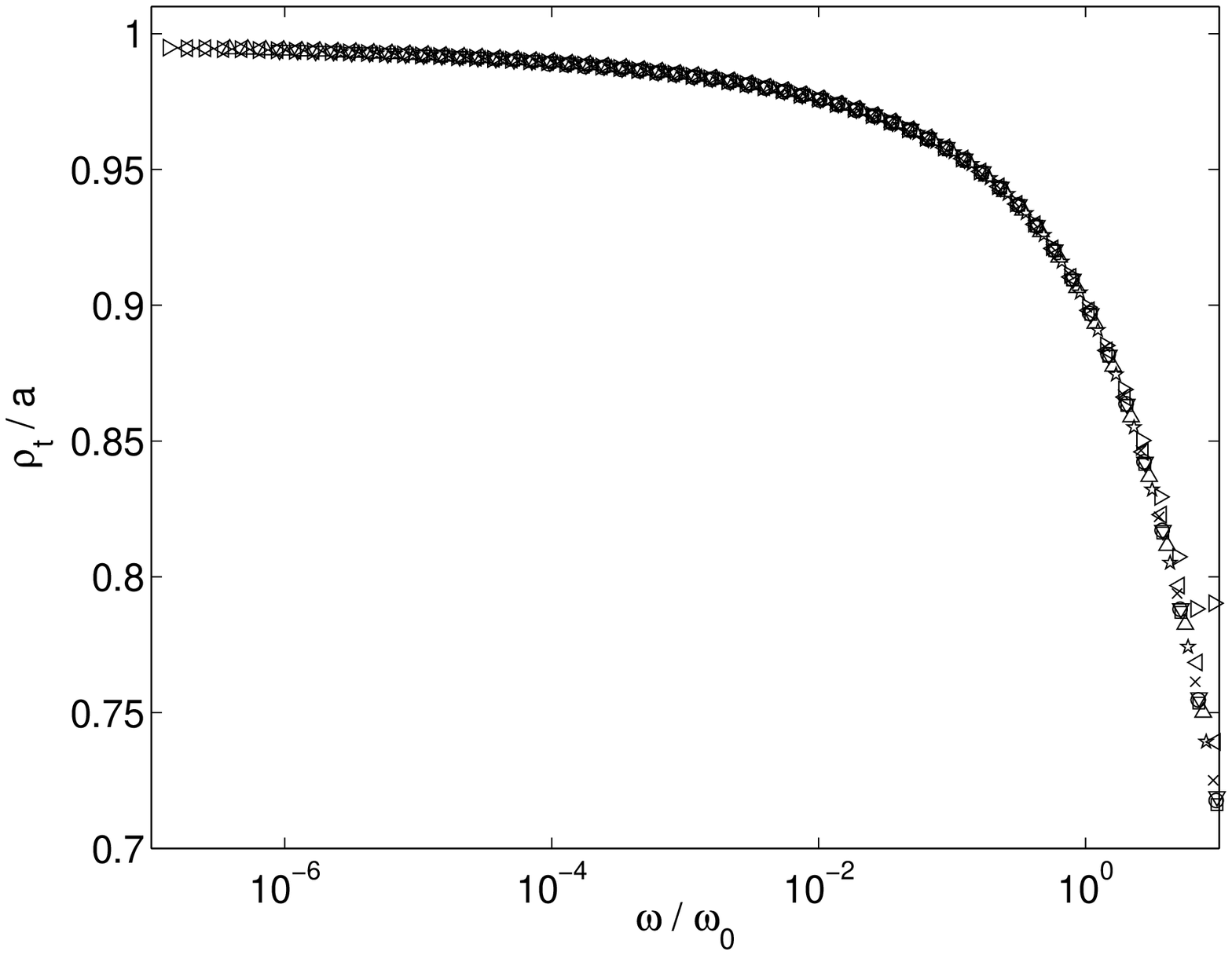}
  \caption{Rescaled spectral densities $\rho_{t}$ of the AFM s-d model with
    $S=1$ (top) for antiferromagnetic couplings $J=0.1
    (\times)$, $J=0.15 (\circ)$, $\ldots$, $J=0.45$.
    One sees that after rescaling the spectrum $\rho$ and
    the frequency $\omega$, all points fall on the same line.
    For better visibility, the inset shows the same data slightly shifted for
    each $J$. Full lines are least square fits to $a - b/\ln^2(T_0/\omega)$.
    The fits have been done on the interval $[10^{-8}, \omega_0]$, where
    $\omega_0$ is taken such that $\rho_t(\omega_0)/\rho_t(0) \approx 0.9$.}
  \label{fig:t_spec_rescale_S2m}
\end{figure}

\begin{figure}
  \includegraphics[width=0.45\textwidth]{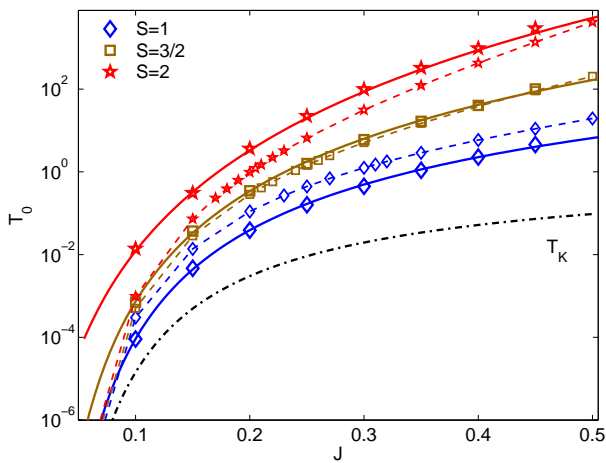}
  \caption{Energy scale $T_0$ as a function of the antiferromagnetic coupling
    $J$ for $S=1$ (diamonds) and $S=3/2$ (squares), and $S=2$ (pentagrams).
    The full lines show fits to
    $T_0(J) = \bar D\,|bJ|^n\,\exp(b/J)$ with different values of $\bar D$,
    $b$ and $n$.
    The dashed lines with the somewhat smaller symbols indicate  $T_0$ as
    extracted from the flow diagrams.
    The dot-dashed line shows $T_K=D\,\sqrt{2 \rho_0 J}\exp(-1/(2 \rho_0 J))$
    with $D=1$ and $\rho_0=0.5$. }
  \label{fig:T0_fit_Sxm}
\end{figure}

Figure~\ref{fig:t_spec_Sx_Jm50} shows $\rho_t(\omega)$ for various values of $S$
and a fixed $J$. The band edge features become more pronounced with increasing
$S$, and $\rho_t(0)$ is independent of $S$. A magnified picture of the
low-energy behaviour is shown in the inset. On this scale, one clearly sees
the qualitative difference between the fully screened $S=1/2$ case (quadratic
behaviour) and the underscreened $S>1/2$ case (cusps).
In the latter case, a good fit to the low-energy behaviour of the spectra
$\rho_t(\omega)$ can be obtained with
\begin{equation}                                    \label{eq:rho_t_asympt}
  \rho_t(\omega) \sim a - \frac{b}{\ln^2(\omega/T_0)}\:.
\end{equation}
The parameter $a$ is independent of $S$ and approximately given by
$a=1/(J^2 \pi^2 \rho_0)$ as inferred from \Eq{eq:rho_d0t} for a vanishing
$\rho_d(0)$.
In Fig.~\ref{fig:t_spec_rescale_S2m} we plot the rescaled spectra on
a logarithmic scale.
On low energy scales, all spectra collapse onto a single curve showing (i) the
validity of \Eq{eq:rho_t_asympt}, and (ii) that $b$ is an implicit function of
$T_0$.
The resulting energy scale $T_0$ is shown in Fig.~\ref{fig:T0_fit_Sxm} as a
function of $J$ for various values of $S$. 
The dashed lines indicate $T_0$ as obtained from the flow diagrams. We observe
that the agreement of the two approaches to extract $T_0$ is not as good as in
the ferromagnetic case.

The reason for this disagreement is that these two $T_0$'s are obtained from
the behaviour on low but rather different energy scales.
There is no {\it a  priori} reason for them to coincide.
\footnote{We could calculate the spectra down to $\omega=10^{-8}$ but the
flows to $10^{-60}$. Therefore, 
the fitting of the Green's functions is done over an energy range of
$[10^{-8}, 10^{-5}]$ and that for the flows on $[10^{-60}, 10^{-40}]$. }
The antiferromagnetic case is more complex than the ferromagnetic.
A higher energy scale (Kondo temperature $T_K$, independent of $S$) is
associated with the partial screening of the original spin.
By contrast, on the lowest energies the residual ferromagnetic
coupling leads to another energy scale $T_0$ depending on $S$, as discussed in
Sec.~\ref{sec:FixedPoint}. A reasonable explanation of the fact that $T_0$
obtained from the spectra differs from the one from the fixed-point analysis
is that the former is still affected by the crossover between these regions.
\section{Scattering Cross Sections}                   \label{sec:scattering}
\begin{figure}
  \includegraphics[width=0.48\textwidth]{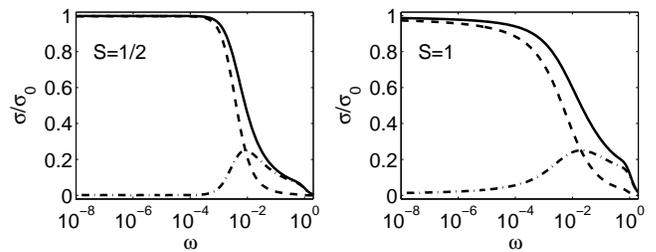}
  \caption{Elastic (dashed), inelastic (dot-dashed) and total (full)
    cross section for the fully screened  model (left) and
    underscreened model ($S=1$, right) for $J=0.2$.} 
  \label{fig:sigma_AFM_J40_Sx}
\end{figure}
\begin{figure}
  \includegraphics[width=0.48\textwidth]{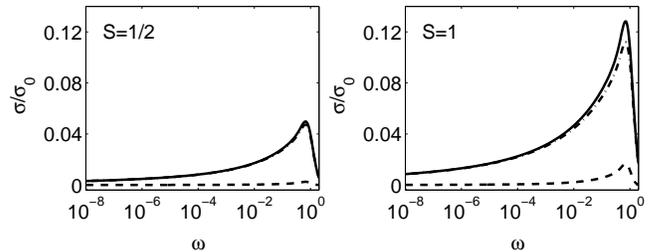}
  \caption{Elastic (dashed), inelastic (dot-dashed) and total (full)
  cross section for the ferromagnetic model $J=-0.2$ The left-hand
  (right-hand) plot is for $S=1/2\:(S=1)$. The dot-dashed and the full
  lines virtually lie on top of each other indicating that the
  scattering is predominantly inelastic.}
  \label{fig:sigma_FM_J40_Sx}
\end{figure}

The on-shell t-matrix $t_{k,k'}$ directly yields the elastic cross section
and can be related to the total scattering cross section via the optical
theorem.
Following Zar\'and {\it et al}\:\cite{ZBDA04pre}, we can write the scattering
cross sections as 
\begin{equation}
  \begin{aligned}
    \sigma_{\rm tot}(k) &= -\frac{2}{v_F}\,\Im t_{k,k}\\
    \sigma_{\rm el}(k)  &= \frac{1}{v_F}\,\int
    \frac{d^3k'}{(2\pi)^3}\,2\pi\delta(\eps_{k'} - \eps_{k})\,|t_{k',k}|^2\:,
  \end{aligned}
\end{equation}
where $v_F$ denotes the Fermi velocity.
The inelastic cross section is then given by the difference
$\sigma_{\rm inel}(k)=\sigma_{\rm  tot}(k)-\sigma_{\rm el}(k)$.
By evaluating these equations for a flat density of states $\rho_0$ and 
$t_{k',k} = J^2\,\alpha_k\,\alpha_{k'}\,G_t(\eps_k)$ with $\alpha_k$ depending
on $k$ only through $\eps_k$, we obtain
\begin{equation}
  \begin{aligned}
    \sigma_{\rm tot}(\omega) &= -\frac{2 J^2\,|\alpha_k|^2}{v_F}\,\Im G_t(\omega)\\
    \sigma_{\rm el}(\omega)  &= \frac{2\pi\rho_0\,N}{v_F}\,J^4\,
    |\alpha_k|^4\,|G_t(\omega)|^2\:.
  \end{aligned}
\end{equation}
This simplifies further if we assume purely isotropic scattering $\alpha_k^2 =
1/N$, and we find
\begin{align}
  \sigma_{\rm tot}(\omega) &= -\sigma_0\rho_0\pi\,J^2\,\Im G_t(\omega)\\
  \sigma_{\rm el}(\omega)  &= \sigma_0\rho_0^2\pi^2\, J^4\,|G_t(\omega)|^2
\end{align}
with $\sigma_0^{-1} = \pi \rho_0 v_F N/2$.
These cross sections are shown in Fig.~\ref{fig:sigma_AFM_J40_Sx} for an
antiferromagnetic coupling and in  Fig.~\ref{fig:sigma_FM_J40_Sx} for the
ferromagnetic case.

In the antiferromagnetic case, the asymptotic behaviour of
$G_t(\omega)$ as $\omega \to 0$ is
$G_t(\omega) \to -i/(\rho_0 \pi J^2)$.
Therefore, on the Fermi level
\begin{equation}
  \sigma_{\rm tot}(0) = \sigma_{\rm el}(0) = \frac{2}{\pi \rho_0 v_F N} \equiv \sigma_0\:.
\end{equation}
This shows that the inelastic scattering cross section vanishes at the
Fermi level, as it should for a Fermi liquid. For the regular Fermi liquid
($S=1/2$, Kondo model), the inelastic scattering vanishes quadratically, and
therfore on the lowest energy scale
$\sigma_{\rm tot}(\omega)=\sigma_{\rm  el}(\omega)$. In contrast to that, for
the underscreened model ($S=1$), the inelastic scattering falls off much more
slowly. 

In the ferromagnetic case, we find that $G_t(\omega) \to 0$ as
$\omega \to 0$, which implies that the total cross section and thus all
scattering vanishes at the Fermi level. Moreover, since
$\sigma_{\rm  el}(\omega)$ involves the square of $G_t(\omega) \ll 1$
for $\omega\to 0$ we expect
$\sigma_{\rm el}(\omega)\ll \sigma_{\rm  tot}(\omega)\sim G_t(\omega)$ which
is a formal explanation of the predominance of inelastic scattering in this case.

\section{Spin Susceptibilities}                             \label{sec:spinsus}
\begin{figure}
  \includegraphics[width=0.47\textwidth]{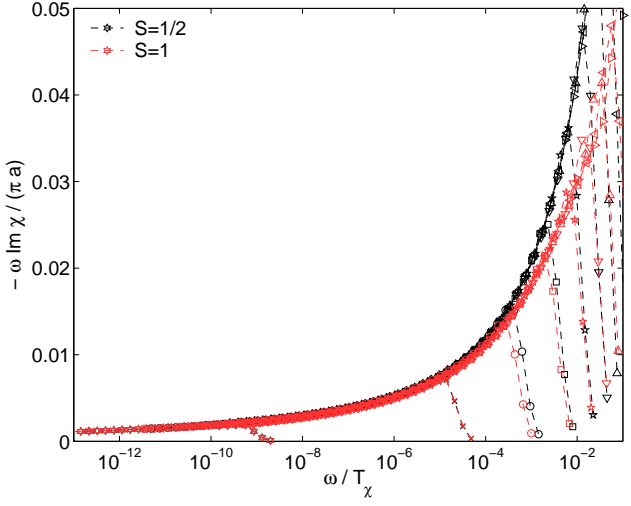}\\[1ex]
  \caption{Low-energy spectra of the impurity spin susceptibility multiplied by
  $\omega$ for  ferromagnetic coupling to $S=1/2$ (dark symbols) and $S=1$ (light
  symbols). The energies are rescaled by $T_\chi$ which is fitted from
  $-\omega\,\Im \chi(\omega)/\pi = a/\ln^2(\omega/T_\chi)$ with a
  logarithmic least squares fit.  Eq.~\ref{eq:chi_fit}. The ranges of
  the fits are $[10^{-8} : 2\times 10^{-2}]$ for $S=1/2$ and $[10^{-8} :
  2\times 10^{-3}]$ for $S=1$. $J$
  ranges from $J=-0.05 (\ast), -0.1 (\times),\ldots$ to $J=-0.45 (>)$.}
  \label{fig:sc_fit_Sxp}
\end{figure}
\begin{figure}
  \includegraphics[width=0.48\textwidth]{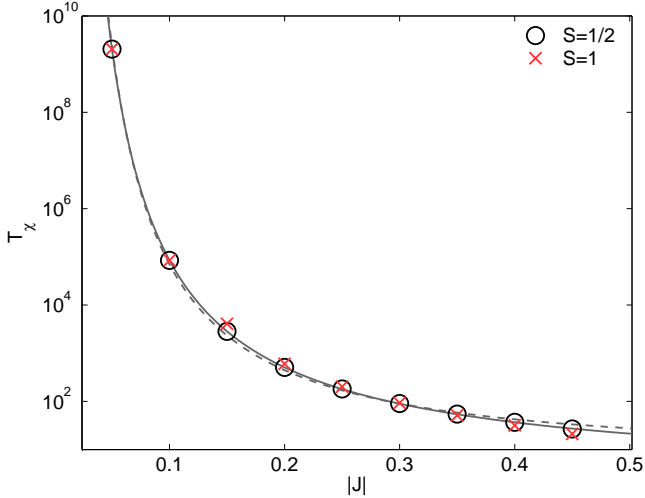}
  \caption{Energy scale $T_\chi$ as a function of $J$ for ferromagnetic
  coupling of a spin $S=1/2 (\circ)$ and  $S=1 (\times)$.
  The broken line is a fit to Eq.~(\ref{eq:Tchi}) with $a=4.7$ and
  $b=0.46$. The full line is a better fit to Eq.~(\ref{eq:Tchi}) with
  $(bJ)^{-1/5}$ instead of the square root and $D=2.55$ and $b=0.50$.}
  \label{fig:Tchi_vs_J_Sxp}
\end{figure}

From the NRG we also calculate the impurity spin susceptilibities
$\chi(\omega)$ both for ferromagnetic and antiferromagnetic coupling.
The low energy behaviour of $\chi(\omega)$ can be obtained from the equations
of motion.
As is shown in appendix \ref{app:spinsus}, the low-energy behaviour is also
singular.
It goes asymptotically for $\omega\to 0$ as
\begin{align}
  -&\frac {1}{\pi} {\rm Im}
  \langle\langle S_z : S_z \rangle\rangle =
  \frac {S(S+1)} {3 \omega \ln^2 (\omega/T_{\chi})}\\
  &\quad\times                              \nonumber
  \left( 1 - \frac{\ln(\ln(\omega/T_{\chi}))}
       {\ln(\omega/T_{\chi})} + {\rm O}
       \left(\frac{(\ln(\ln(\omega/T_{\chi}))^2}
            {\ln^2(\omega/T_{\chi})}
            \right)\right)\:,
\end{align}
where we have introduced an  energy scale $T_\chi$ which we expect to vary as
\begin{equation}                                           \label{eq:Tchi}
  T_\chi = D\,\sqrt{2 \rho_0 |J|}\,e^{-1/(2 J \rho_0)}\:.
\end{equation}
This leads to the following ansatz for the low-energy behaviour of the
spectrum of $\chi(\omega)$:
\begin{equation}                                        \label{eq:chi_fit}
  -\frac{1}{\pi}\,\Im \chi(\omega) = \frac{a} {\omega \ln^2(\omega/T_\chi)}
\end{equation} 
\begin{figure}
  \includegraphics[width=0.47\textwidth]{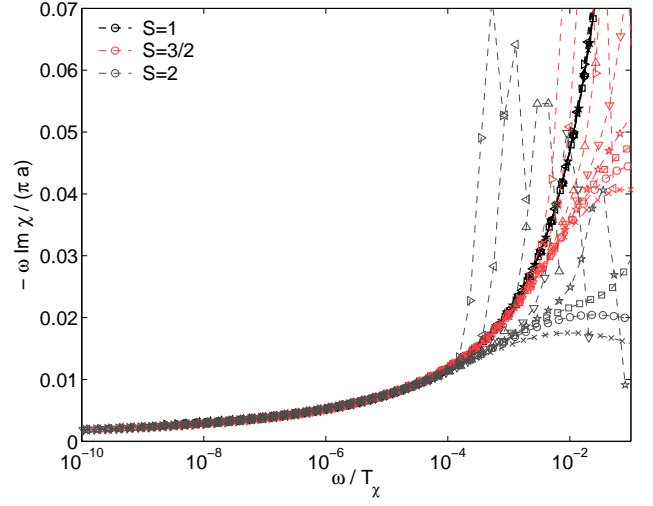}\\[1ex]
  \caption{Low-energy spectra of the impurity spin susceptibility multiplied by
  $\omega$ for antiferromagnetic underscreened models $S=1$ (dark symbols) and
  $S=3/2$ (light red symbols) and $S=2$ (light grey symbols).
  The energies are rescaled by $T_\chi$ which is fitted from
  $-\omega \,\Im \chi(\omega) /\pi = a/\ln^2(\omega/T_\chi)$ with a
  logarithmic least squares fit \Eq{eq:chi_fit}. The ranges of
  the fits vary significantly as a function of $J$ to fit the low-energy behaviour.
  $J$ ranges from $J=0.05 (\ast), 0.1 (\times),\ldots$ to $J=0.45 (>)$. }
  \label{fig:sc_fit_Sxm}
\end{figure}
\begin{figure}
  \includegraphics[width=0.48\textwidth]{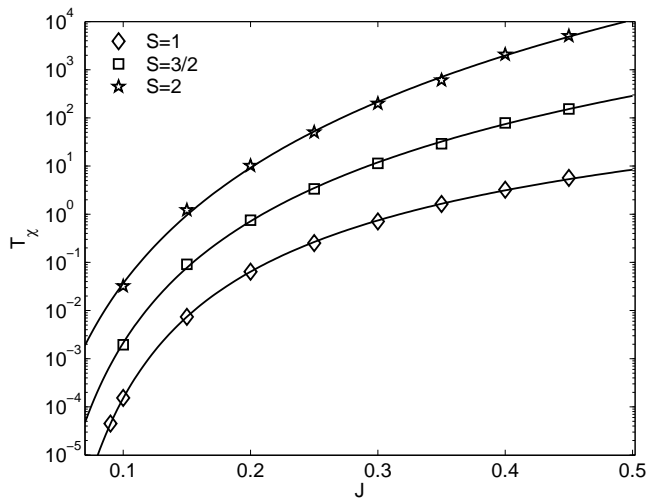}
  \caption{Energy scale $T_\chi$ for underscreened s-d models with
  $S=1\, (\diamond)$, $S=3/2\, (\square)$, and $S=2\, (\star)$. The values
  of $T_\chi$ are obtained from low-energy fits to \Eq{eq:chi_fit}.
  Full lines are fits to \Eq{eq:Tchi} with the square root replaced by the
  exponent $n=2.5$ for  $S=1$,  $n=5.0$ for $S=3/2$, and $n=7.5$ for  $S=2$.}
  \label{fig:T_chi_Sxm}
\end{figure}

As can be seen from Figs.~\ref{fig:sc_fit_Sxp} and \ref{fig:sc_fit_Sxm}, this
formula describes the low-energy susceptibilities in both the ferromagnetic
and the antiferromagnetic cases very well.
In the ferromagnetic case, $a\approx S(S+1)/3$ very well with an error of less
than $1\%$. In the antiferromagnetic case, $S$ needs to be replaced
by $S-1/2$. However, here the agreement is less accurate with an error of the
order of $10 \%$.

The characteristic energy scale $T_\chi$ is plotted in
Figs.~\ref{fig:Tchi_vs_J_Sxp} and \ref{fig:T_chi_Sxm}.
We find that $T_\chi \approx 2\times T_0$ both for ferromagnetic coupling and also
for antiferromagnetic coupling, with $T_0$ from the fitting of the Green's functions
over the same energy scale.

\section{Conclusions}                                     \label{sec:concl}

Our NRG calculations give a comprehensive picture of the dynamics of
the underscreened and the ferromagnetically coupled impurity s-d
models. In the ferromagnetic case, the ground state has $(2S+1)$-fold
degeneracy as the impurity spin becomes completely decoupled.
The NRG excitations, however, only slowly approach their fixed
point values  as $1/(N+C(J))$ as $N\to\infty$. This behaviour  is
consistent with a frequency dependent renormalized exchange coupling
$\tilde J(\omega)=1/\ln(\omega/T_0)$, which is a marginally irrelevant 
operator at this fixed point.
The  energy scale $T_0$, which we calculated in terms of the bare parameters
was found to be independent of the spin $S$.
Its value, as well as the $T_\chi$ deduced from the dynamic susceptibility,
corresponds to the usual expression for $T_K$, but as $J<0$, it is such that
$T_K \gg D$.

Essentially the same kind of behaviour was found in the case of the
underscreened antiferromagnetic model. After the initial screening of
a component of the spin by the conduction electrons in the single
channel, the effective spin value is reduced to $S-1/2$. The effective
exchange coupling $\tilde J(\omega)$ on the low energy scale was also
found to be of the form, $\tilde J(\omega)=1/\ln(\omega/T_0)$, and
hence is ferromagnetic for $\omega<T_0$. 
The values of $T_0$ obtained in this case, however, were found to
depend on the spin value $S$, and to be consistently smaller than in
the ferromagnetic case,  such that $T_0\sim D$.
The value of $T_0$ does not coincide with the formula for the
antiferromagnetic $T_K$ which is independent of $S$ and always smaller than
our values for $T_0$. The latter, however, does agree well with the energy
scale $T_\chi$ deduced from the dynamic susceptibility.

For large values of the bare antiferromagnetic interaction
$J$ we were able to estimate the value an effective bare ferromagnetic
interaction $J_{\rm eff}$ which would lead to the calculated values of
$T_0$.
From the NRG energy levels in the approach to the fixed point, the
asymptotic form of the phase shifts as $\omega\to 0$ due to the
elastic scattering of a single particle excitation in the two
channels $S_r\pm 1/2$ by  the residual spin $S_r$ could be estimated,
where $S_r=S$ in the ferromagnetic case and $S_r=S-1/2$ for the
antiferromagnetic. These were found to be in complete agreement with
the values from an explicit calculation of the scattering 
using the effective low energy Hamiltonian (see Appendix~\ref{app:exscatt}).

The singular form of the effective coupling $\tilde J(\omega)$ is
manifest as sharp cusps at the Fermi level in the spectral
densities calculated for the Green's functions $G_d(\omega)$ and
$G_t(\omega)$. For antiferromagnetic coupling the value of
$\rho_t(\omega)$ at $\omega=0$ was found not to diverge and to 
approach the value $1/\pi^2\rho_0J^2$ independent of the spin
value $S$, as conjectured in~\Eq{eq:rho_of_zero}.
For ferromagnetic coupling the value of $\rho_t(0)$ was
found to be zero for finite $J$ but equal to $S(S+1)\rho_0$ for
$J=0$. This clearly demonstrates that $J=0$ is a singular point, and
the behaviour is discontinuous as a function of $J$ when $J\to 0$ when
approached from the ferromagnetic as well as from the
antiferromagnetic side. The approach of $\rho_d(\omega)$ 
and $\rho_t(\omega)$ to their values at $\omega=0$ could be described
well by terms of the form $1/\ln^2(\omega/T_0)$ in all cases where
there is incomplete screening of the local spin. The energy scale deduced
from this fitting broadly agrees with the one from the fixed-point
analysis. In the case of ferromagnetic coupling the agreement is rather
precise.

The results of the calculations for the inelastic and elastic
scattering cross-sections as a function of frequency $\omega$ are
consistent with the classification of the low energy  behaviour
of these systems with unscreened residual spins as singular Fermi
liquids, as proposed by Mehta et al \cite{MBZAC04pre}. The inelastic
scattering goes to zero as $\omega\to 0$ as to be expected 
for a Fermi liquid, in contrast to non-Fermi liquid behaviour
which is characterized by finite inelastic scattering at $\omega=0$.
The approach  of the inelastic scattering component to zero in the
underscreened cases, however, was found to be anomalously slow compared
to the corresponding results for the  fully screened antiferromagnetic
model with $S=1/2$.  The singular nature of the low energy
scattering is also evident in the contrasting results for the spectral
densities of $\rho_t(\omega)$ as $\omega\to 0$ in the
antiferromagnetic case for $S=1/2$ and $S>1/2$, shown in the inset of
Fig.~\ref{fig:t_spec_Sx_Jm50}.
In the case of ferromagnetic coupling, inelastic scattering  gives the
dominant contribution to the total scattering cross section.

To summarize: our results are in broad agreement with the conclusions
of previous studies of the dynamics of underscreened s-d models
\cite{PG97,CP03,MBZAC04pre,CP04pre}. However, as well as fully analysing the
fixed point behaviour we have  been able to make precise predictions
for the dynamics of a range  of physical response functions, and
have  also been able to calculate the explicitly the renormalized
energy scales $T_0$, $T_\chi$ etc in terms of the parameters of the
bare model.

\begin{acknowledgments}

We wish to thank the EPSRC (Grant GR/S18571/01) for financial
support.
This work was partially supported by SunnyNames llp.

\end{acknowledgments}

\appendix

\section{Exchange scattering in the one-electron case}           \label{app:exscatt}

Consider a system of free electrons that couple to a spin $S$ with a local
exchange scattering term. The Hamiltonian is given by the sum of \Eq{eq:Hband}
and  \Eq{eq:Hint}.
We calculate the one-electron states using the basis set
$\{\ket{k,\sigma: S,M_S} =\cdag_{k,\sigma} \ket{0\rangle |S,M_S}\}$.
The Hamiltonian acts on these states as
\begin{equation}
  \begin{aligned}
    H &\ket{k,\uparrow :S,M_S} =\epsilon_k \ket{k,\uparrow:S,M_S}
    +\alpha_k J\sum_{k'}\alpha_{k'} \\
    &\times \Big(\sqrt{(S-M_S)(S+M_S+1)}\, \ket{k',\downarrow :S,M_S+1}\\ 
    &+ M_S \,\ket{k',\uparrow :S,M_s}\Big)\:.
  \end{aligned}
\end{equation}
and
\begin{align}
    H & \ket{k,\downarrow :S,M_S+1} = \epsilon_k \ket{k,\downarrow:S,M_S+1}
    +\alpha_k J\sum_{k'}\alpha_{k'} \nonumber \\
    &\times \Big(\sqrt{(S-M_S)(S+M_S+1)}\,\ket{k',\uparrow :S,M_S} \nonumber\\
    &-  (M_S+1)\,\ket{k',\downarrow :S,M_S+1}\Big)\:.
\end{align}
Solving the eigenvalue equation we find the $M_S$ dependence cancels out, and
the equation factorizes so that we get two independent equations for the new
energy levels. 
With the free $d$-site green's function  as defined in \Eq{eq:G_d0}, we obtain
\begin{equation}
  1 = JS\,G_d^{(0)}(\omega)
\end{equation}
which corresponds to scattered states in the $S+1/2$ channel and
\begin{equation}
  1 = -J(S+1)\,G_d^{(0)}(\omega)
\end{equation}
which correspond to scattering in the $S-1/2$ channel.

The one electron eigenstates in $S+1/2$ channel are built up from the states, 
\begin{equation}
  \begin{aligned} 
    \ket{k,M_S+1/2} &=\left({S+M_S+1\over 2S+1}\right)^{1/2}
    \ket{k,\uparrow :S,M_S}  \\
    &+\left({S-M_S\over 2S+1}\right)^{1/2}\ket{k,\downarrow :S, M_S+1}
  \end{aligned} 
\end{equation}
and those for the $S-1/2$ channel  from the states,
\begin{equation}
  \begin{aligned} 
    \ket{k,M_S+1/2} &=-\left({S-M_S\over 2S+1}\right)^{1/2}
    \ket{k,\uparrow :S, M_S}\\
    &+\left({S+M_S+1\over 2S+1}\right)^{1/2}\ket{k,\downarrow :S, M_S+1}
  \end{aligned}
\end{equation}
We can define corresponding creation operators, $A_{k,M_S+1/2}^{\dagger}$ via 
\begin{align}                                              \label{eq:def_A}
  A_{k,M_S+1/2}^{\dagger}&=\sqrt{{S+M_S+1\over 2S+1}}
  f(S,M_S)(S^+)^{(M_S+S)}c_{k,\uparrow}^{\dagger} \\
  &+\sqrt{{S-M_S\over 2S+1}}f(S, M_S+1)(S^+)^{(M_S+S+1)}c_{k,\downarrow}\nonumber
\end{align}
where $M_S$ can only take values $S, S-1, \ldots, -S-1$ as $(S^+)^{2S+1}=0$,
and $B_{k,M_S+1/2}^{\dagger}$ via
\begin{align}                                               \label{eq:def_B}
  B_{k,M_S+1/2}^{\dagger} &= -\sqrt{{S\!-\!M_S\over 2S+1}}f(S,
  M_S)(S^+)^{(M_S+S)}c_{k,\uparrow}^{\dagger}\\
  &+\sqrt{{S+M_S+1\over 2S+1}}f(S, M_S+1)(S^+)^{(M_S+S+1)}c_{k,\downarrow}\nonumber
\end{align}
where $M_S$ can take values $S-1, \ldots, -S$.
These operators operate on the vacuum state $\ket{0\rangle|S,-S}$,
and $f(S,M_S)$ is a normalization factor such that
$f(S,M_S)(S^+)^{(M_S+S)} \ket{S,-S}=\ket{S,M_S}$.

Confining our attention to the system with no more than one electron we find 
equations for the Green's functions
$G^A_{k,k',M_S+1/2}(\omega) =\biggreen{A_{k,M_S+1/2}^{}:
  A_{k',M_S+1/2}^{\dagger}}$
and 
$G^B_{k,k',M_S+1/2}(\omega)=\biggreen{B_{k,M_S+1/2}^{}:
  B_{k',M_S+1/2}^{\dagger}}$ as
\begin{align}
  (\omega-\epsilon_k)\,&G^A_{k,k',M_S+1/2}(\omega) = \delta_{k,k'} \\
  &+\alpha_kJS\sum_{k''}\alpha_{k''}G^A_{k'',k',M_S+1/2}(\omega) \nonumber
\end{align}
and
\begin{align}
  (\omega-\epsilon_k)\,&G^B_{k,k',M_S+1/2}(\omega) = \delta_{k,k'} \\
  -&\alpha_kJ(S+1)\sum_{k''}\alpha_{k''}G^B_{k'',k',M_S+1/2}(\omega) \nonumber\:.
\end{align} 
Solving these equations we obtain
\begin{align}
  G^A_{k,k',M_S+1/2}(\omega) &= {\delta_{k,k'} \over \omega-\epsilon_{k'}}\\
  &+{\alpha_k\over \omega-\epsilon_k}{JS\over1-JSG_d^{(0)}(\omega)}
  {\alpha_{k'}\over \omega-\epsilon_{k'}} \nonumber
\end{align} 
and
\begin{align}
  G^B_{k,k',M_S+1/2}(\omega) &= {\delta_{k,k'}\over \omega-\epsilon_{k'}}\\
  &- {\alpha_k\over \omega-\epsilon_k}{J(S+1)\over1+J(S+1)G_d^{(0)}(\omega)}
  {\alpha_{k'}\over \omega-\epsilon_{k'}} \nonumber\:.
\end{align}

This enables us to calculate also the Green's function $G_t(\omega)$ as
defined in \Eq{eq:def_Gt} for the one-electron case.
In terms of the creation and annihilation operators defined in equations
(\ref{eq:def_A}) and (\ref{eq:def_B}), we find
\begin{align}
  S^+\ddag_{\downarrow} + S_z \ddag_{\uparrow} =&
  \sum_k{\alpha_k\over\sqrt{2S+1}} \\
  \times& \Big\{S \sqrt{S+M_S+1} A_{k,M_S+1/2}^{\dagger}    \nonumber \\
  +&(S+1)\sqrt{S-M_S} B_{k,M_S+1/2}^{\dagger}\Big\}\:.      \nonumber
\end{align}
The Green's function $G_t(\omega)$ for this one electron situation
is then given by
\begin{align}                                           \label{eq:tgf}
  G_t(\omega) &= \sum_{k,k',M_S}{\alpha_k\alpha_{k'}\over 2(2S+1)^2}\\
  &\Big\{S^2(S+M_S+1)\,G^A_{k,k',M_S+1/2}(\omega)  \nonumber\\
  &+(S+1)^2(S-M_S)\,G^B_{k,k',M_S+1/2}(\omega)\Big\} \nonumber\:.
\end{align}
We find
\begin{equation}
  \sum_{k,k'}\alpha_k\alpha_{k'}G^A_{k,k',M_S+1/2}(\omega) =
      {G_d^{(0)}(\omega) \over 1-JSG_d^{(0)}(\omega) }
\end{equation}
and
\begin{equation}
  \sum_{k,k'}\alpha_k\alpha_{k'}G^B_{k,k',M_S+1/2}(\omega)=
      {G_d^{(0)}(\omega) \over 1+J(S+1)G_d^{(0)}(\omega) }\:.
\end{equation}
We substitute these into \Eq{eq:tgf} and perform the sums over $M_S$.
This sum has to be done carefully because the for the first term the sum is
over the values $-S-1, -S, \ldots, S$, whereas for the second term it is over
$-S, -S+1, \ldots, S-1$. 
The result is 
\begin{align}
  G_t(\omega)&={S(S+1)G_d^{(0)}(\omega) \over 2} \\
  &\times \left\{{1\over 1-JSG_d^{(0)}(\omega) } +
         {1\over 1+J(S+1)G_d^{(0)}(\omega) }\right\}   \nonumber \:.
\end{align}
In our many-electron case this theory is only directly applicable in the
immediate vicinity of the fixed point with one quasiparticle excited from the
interacting ground state and $J$ interpreted as the renormalized coupling
$\tilde J$.
The phase shifts for the quasiparticle scattering are then given by
\begin{equation}                                             \label{eq:nps}
  \begin{aligned}
    \eta_{S+1/2}(\omega) &=-\arctan\big(\pi\rho_0(\omega)\tilde J(\omega)S\big)\:,\\
    \eta_{S-1/2}(\omega) &=\quad\arctan\big(\pi\rho_0(\omega)\tilde J(\omega)(S+1)\big)
  \end{aligned}
\end{equation} 
with $\tilde J(\omega) = 1/\ln(\omega/T_0)$.
In the limit $\omega \to 0$, 
the phase shifts we calculate in this way agree with the estimates
\Eq{eq:phaseshift1_FM} and \Eq{eq:phaseshift2_FM} obtained from the NRG
flows. 

\section{Green's Functions and t-Matrices for the Anderson Model} \label{app:Anderson}

Let the impurity site be labelled by $f$ and next site it is coupled to on the
chain be labelled $d$, as indicated in Fig.~\ref{fig:geometry}.
Let $d_\sigma = \sum_{k} V_k c_k/V$ where $|V|^2=\sum_k|V_k|^2$. 
The equation of motion for the Green's function $G_{kk'}(\omega)$ yields the
relation 
\begin{equation}
  G_{kk'}(\omega)={\delta_{kk'}\over \omega -\epsilon_k}+
  {V\alpha_{k}\over \omega-\epsilon_k}\,G_f (\omega)\,
  {V\alpha_{k'}\over \omega -\epsilon_k'}\:.
\end{equation}
The t-matrix $t_{kk'}(\omega)$ is defined by the equation,
\begin{equation}
  G_{kk'}(\omega)={\delta_{kk'}\over \omega -\epsilon_k}+{1\over \omega
  -\epsilon_k}\,t_{kk'}(\omega)\,{1\over \omega -\epsilon_k'}\:,
\end{equation}
and hence for the Anderson model we have
\begin{equation}
  t_{kk'}(\omega)=V^2 \alpha_{k}G_{f}(\omega)\alpha_{k'}\:.
\end{equation}

We can derive an expression for $G_{d}(\omega)$ from $G_{kk'}(\omega)$ by
multiplying by $V_k$, $V_{k'}$, and summing over $k$ and $k'$.
This gives
\begin{equation}
  G_{d}(\omega) = G_d^{(0)}(\omega) +
  V^2 G_d^{(0)}(\omega)\,G_{f}(\omega)\, V^2\, G_d^{(0)}(\omega)\:.
\end{equation}
In the wide band limit $V^2\,G_d^{(0)}(0)=-i\Delta$ and hence,
\begin{equation}
  G_{d}(\omega)={1\over V^2}\,
  \left[-i\Delta-\Delta^2 G_{f}(\omega)\right]\:.
\end{equation}

We can deduce the spectral density of $G_{d}(\omega)$  from
\begin{equation}                               \label{eq:g00AM}
  \rho_{d}(\omega)=-{1\over\pi}{\Im}G_{f}(\omega)=
  \rho_0 [1-\pi\Delta\rho_{f}(\omega)]\:,
\end{equation}
where  we have used the result $\Delta=\pi\rho_0|V|^2$, where $\rho_0$ is the
free conduction density of states at the Fermi level.
It follows from this  expression that for the symmetric model, where
$\rho_f(0)=1/\pi\Delta$, the spectral density of $G_d$ vanishes,
$\rho_{d}(0)=0$.
Therefore, we find an anti-resonance.

\section{Dynamic Spin susceptibility}                   \label{app:spinsus}

We derive equations of motion for the local spin-spin double-time  Green's
function $\langle\langle S^+(t) : S^-(t')\rangle\rangle$ to second order in
the coupling $J$.
For this we need the evaluation of the higher order Green's function to zero
order only. We find from the equations of motion
\begin{equation}
  \omega^2\langle\langle S^+:S^-\rangle\rangle_\omega
  ={8J^2S(S+1)\over 3}\langle\langle S_d^+:S_d^-\rangle\rangle_\omega\:,
\end{equation} 
where we have used the fact the
$\langle\langle S^+:S^-\rangle\rangle=2\langle\langle S_z:S_z\rangle\rangle$,
due to the rotational symmetry in zero magnetic field. 
The spin operator $S_d^+$ for the first site on the chain in terms of the
conduction electron states is given by
$S_d^+= \ddag_\UA \dnod_\DA = 
\sum_{k,k'}\alpha_k\alpha_{k'}\cdag_{k\uparrow}\cnod_{k'\downarrow}$,
with a similar expression for $S_d^-$.
Evaluation of $\langle\langle S_d^+:S_d^-\rangle\rangle$ to zero order in $J$
gives 
\begin{equation}
  \langle\langle S_d^+ : S_d^- \rangle\rangle
  =\int\!\!\!\!\int \rho_0(\epsilon)
  \rho_0(\epsilon')\left({f(\epsilon)-f(\epsilon')\over
  \omega-\epsilon'+\epsilon}\right)\, d\epsilon \,d\epsilon'\:,
\end{equation} 
where $\rho_0(\omega)=\sum_{k}\alpha_k^2\delta(\omega-\epsilon)$ is the
local spectral density for the first site on the chain for the non-interacting
system. From that we deduce that the spectral density of
$\langle\langle S^+ : S^- \rangle\rangle$ to order $J^2$
is given by 
\begin{equation}
  \begin{aligned}
    -{1\over\pi} & {{\rm Im}\langle\langle S^+:S^-\rangle\rangle} 
    = {8J^2S(S+1)\over 3\,\omega^2}\\
    &\times
    \int\rho_0(\epsilon)
    \rho_0(\epsilon+\omega)(f(\epsilon)-f(\epsilon+\omega))\,d\epsilon \:.
  \end{aligned}
\end{equation} 
For the limit $\omega\to 0$ we find 
\begin{equation}
  -{1\over\pi}\lim_{\omega\to 0}\omega\,
  \Im\langle\langle S^+ : S^- \rangle\rangle = 
  {8(J\rho_0(0))^2  S(S+1) \over 3}
\end{equation} 
Using the poor man's scaling equation, Eq. (3.51) in
Ref.~\onlinecite{hewson}, with $\tilde D = a\,\omega$,
the coupling $J$ is renormalized to $\tilde J(\omega)$ given by
\begin{equation}                            \label{eq:scale_J_chi}
  2\tilde J\rho_0 = 
  {1\over \ln(\omega/T_{\chi})} - 
  {\ln(\ln(\omega/T_{\chi})
    \over 2 \ln^2 (\omega/T_{\chi})} +{\rm O}
  \left({1\over \ln^2(\omega/T_{\chi})}\right) \:,
\end{equation}
with the energy scale $T_\chi$ given by
\begin{equation}
  T_{\chi} = D|2J\rho_0|^{1/2}e^{-1/2J\rho_0}\:.
\end{equation}
Hence, asymptotically as $\omega\to 0$,
\begin{equation}
  \begin{aligned}
    -&\frac {1}{\pi} {\rm Im}
    \langle\langle S^+:S^- \rangle\rangle =
    \frac {2 S(S+1)} {3 \omega \ln^2 (\omega/T_{\chi})}\\
    &\times
    \left( 1 - \frac{\ln(\ln(\omega/T_{\chi}))}
         {\ln(\omega/T_{\chi})} + {\rm O}
         \left(\frac{(\ln(\ln(\omega/T_{\chi}))^2}
              {\ln^2(\omega/T_{\chi})}
              \right)\right) \:.
  \end{aligned}
\end{equation}
For the fully screened Kondo model, this asymptotic form should be
appropriate only for the regime $D>\omega>T_\chi=T_K$.
However, for a ferromagnetic coupling and for the underscreened models, this
asymptotic form should also apply as $\omega \to 0$,
as confirmed in Sec.~\ref{sec:spinsus}.
As the asymptotic behaviour of the  underscreened models is associated with a
ferromagnetic fixed point, the low-temperature $T_\chi$ is  expected
to differ from $T_K$ (high temperature), see Sec.~\ref{sec:spinsus}.


\begin{thebibliography}{32}
\expandafter\ifx\csname natexlab\endcsname\relax\def\natexlab#1{#1}\fi
\expandafter\ifx\csname bibnamefont\endcsname\relax
  \def\bibnamefont#1{#1}\fi
\expandafter\ifx\csname bibfnamefont\endcsname\relax
  \def\bibfnamefont#1{#1}\fi
\expandafter\ifx\csname citenamefont\endcsname\relax
  \def\citenamefont#1{#1}\fi
\expandafter\ifx\csname url\endcsname\relax
  \def\url#1{\texttt{#1}}\fi
\expandafter\ifx\csname urlprefix\endcsname\relax\def\urlprefix{URL }\fi
\providecommand{\bibinfo}[2]{#2}
\providecommand{\eprint}[2][]{\url{#2}}

\bibitem[{\citenamefont{Coleman et~al.}(2001)\citenamefont{Coleman, P\'epin,
  Si, and Ramazashvili}}]{CPSR01}
\bibinfo{author}{\bibfnamefont{P.}~\bibnamefont{Coleman}},
  \bibinfo{author}{\bibfnamefont{C.}~\bibnamefont{P\'epin}},
  \bibinfo{author}{\bibfnamefont{Q.}~\bibnamefont{Si}}, \bibnamefont{and}
  \bibinfo{author}{\bibfnamefont{R.}~\bibnamefont{Ramazashvili}},
  \bibinfo{journal}{J. Phys.: Condens. Matter} \textbf{\bibinfo{volume}{13}},
  \bibinfo{pages}{R723} (\bibinfo{year}{2001}).

\bibitem[{\citenamefont{von L\"ohneysen}(1996)}]{Loe96}
\bibinfo{author}{\bibfnamefont{H.}~\bibnamefont{von L\"ohneysen}},
  \bibinfo{journal}{J. Phys.: Condens. Matter} \textbf{\bibinfo{volume}{8}},
  \bibinfo{pages}{9689} (\bibinfo{year}{1996}).

\bibitem[{\citenamefont{Steglich}(2000)}]{Ste00}
\bibinfo{author}{\bibfnamefont{F.}~\bibnamefont{Steglich}},
  \bibinfo{journal}{Phys. Rev. Lett.} \textbf{\bibinfo{volume}{85}},
  \bibinfo{pages}{626} (\bibinfo{year}{2000}).

\bibitem[{\citenamefont{Hertz}(1976)}]{Her76}
\bibinfo{author}{\bibfnamefont{J.}~\bibnamefont{Hertz}},
  \bibinfo{journal}{Phys. Rev. B} \textbf{\bibinfo{volume}{14}},
  \bibinfo{pages}{1165} (\bibinfo{year}{1976}).

\bibitem[{\citenamefont{Millis}(1993)}]{Mil93}
\bibinfo{author}{\bibfnamefont{A.~J.} \bibnamefont{Millis}},
  \bibinfo{journal}{Phys. Rev. B} \textbf{\bibinfo{volume}{48}},
  \bibinfo{pages}{7183} (\bibinfo{year}{1993}).

\bibitem[{\citenamefont{Continentino}(1993)}]{Con93}
\bibinfo{author}{\bibfnamefont{M.~A.} \bibnamefont{Continentino}},
  \bibinfo{journal}{Phys. Rev. B} \textbf{\bibinfo{volume}{47}},
  \bibinfo{pages}{11587} (\bibinfo{year}{1993}).

\bibitem[{\citenamefont{Schr\"oder et~al.}(1998)\citenamefont{Schr\"oder,
  Aeppli, Bucher, Ramazashvili, and Coleman}}]{SABRC98}
\bibinfo{author}{\bibfnamefont{A.}~\bibnamefont{Schr\"oder}},
  \bibinfo{author}{\bibfnamefont{G.}~\bibnamefont{Aeppli}},
  \bibinfo{author}{\bibfnamefont{E.}~\bibnamefont{Bucher}},
  \bibinfo{author}{\bibfnamefont{R.}~\bibnamefont{Ramazashvili}},
  \bibnamefont{and} \bibinfo{author}{\bibfnamefont{P.}~\bibnamefont{Coleman}},
  \bibinfo{journal}{Phys. Rev. Lett.} \textbf{\bibinfo{volume}{80}},
  \bibinfo{pages}{5623} (\bibinfo{year}{1998}).

\bibitem[{\citenamefont{Bulla et~al.}(1997)\citenamefont{Bulla, Pruschke, and
  Hewson}}]{BPH97}
\bibinfo{author}{\bibfnamefont{R.}~\bibnamefont{Bulla}},
  \bibinfo{author}{\bibfnamefont{T.}~\bibnamefont{Pruschke}}, \bibnamefont{and}
  \bibinfo{author}{\bibfnamefont{A.~C.} \bibnamefont{Hewson}},
  \bibinfo{journal}{J. Phys.: Condens. Matter} \textbf{\bibinfo{volume}{9}},
  \bibinfo{pages}{10463} (\bibinfo{year}{1997}).

\bibitem[{\citenamefont{Gonzalez-Buxton and Ingersent}(1998)}]{GI98}
\bibinfo{author}{\bibfnamefont{C.}~\bibnamefont{Gonzalez-Buxton}}
  \bibnamefont{and}
  \bibinfo{author}{\bibfnamefont{K.}~\bibnamefont{Ingersent}},
  \bibinfo{journal}{Phys. Rev. B} \textbf{\bibinfo{volume}{57}},
  \bibinfo{pages}{14254} (\bibinfo{year}{1998}).

\bibitem[{\citenamefont{Ingersent and Si}(2002)}]{IS02}
\bibinfo{author}{\bibfnamefont{K.}~\bibnamefont{Ingersent}} \bibnamefont{and}
  \bibinfo{author}{\bibfnamefont{Q.}~\bibnamefont{Si}}, \bibinfo{journal}{Phys.
  Rev. Lett.} \textbf{\bibinfo{volume}{89}}, \bibinfo{pages}{076403}
  (\bibinfo{year}{2002}).

\bibitem[{\citenamefont{Glossop and Logan}(2003)}]{GL03}
\bibinfo{author}{\bibfnamefont{M.~T.} \bibnamefont{Glossop}} \bibnamefont{and}
  \bibinfo{author}{\bibfnamefont{D.~E.} \bibnamefont{Logan}},
  \bibinfo{journal}{J. Phys.: Condens. Matter} \textbf{\bibinfo{volume}{15}},
  \bibinfo{pages}{7519} (\bibinfo{year}{2003}).

\bibitem[{\citenamefont{Fritz and Vojta}(2004)}]{FV04pre}
\bibinfo{author}{\bibfnamefont{L.}~\bibnamefont{Fritz}} \bibnamefont{and}
  \bibinfo{author}{\bibfnamefont{M.}~\bibnamefont{Vojta}}
  (\bibinfo{year}{2004}), \bibinfo{note}{cond-mat/0408543}.

\bibitem[{\citenamefont{Coleman and P\'epin}(2003)}]{CP03}
\bibinfo{author}{\bibfnamefont{P.}~\bibnamefont{Coleman}} \bibnamefont{and}
  \bibinfo{author}{\bibfnamefont{C.}~\bibnamefont{P\'epin}},
  \bibinfo{journal}{Phys. Rev. B}  (\bibinfo{year}{2003}).

\bibitem[{\citenamefont{Fateev and Wiegmann}(1981)}]{FW81}
\bibinfo{author}{\bibfnamefont{V.~A.} \bibnamefont{Fateev}} \bibnamefont{and}
  \bibinfo{author}{\bibfnamefont{P.~B.} \bibnamefont{Wiegmann}},
  \bibinfo{journal}{Phys. Lett.} \textbf{\bibinfo{volume}{81A}},
  \bibinfo{pages}{179} (\bibinfo{year}{1981}).

\bibitem[{\citenamefont{Tsvelick and Wiegmann}(1983)}]{TW83}
\bibinfo{author}{\bibfnamefont{A.~M.} \bibnamefont{Tsvelick}} \bibnamefont{and}
  \bibinfo{author}{\bibfnamefont{P.~B.} \bibnamefont{Wiegmann}},
  \bibinfo{journal}{Adv. Phys.} \textbf{\bibinfo{volume}{32}},
  \bibinfo{pages}{453} (\bibinfo{year}{1983}).

\bibitem[{\citenamefont{Andrei et~al.}(1983)\citenamefont{Andrei, Furuya, and
  Lowenstein}}]{AFL83}
\bibinfo{author}{\bibfnamefont{N.}~\bibnamefont{Andrei}},
  \bibinfo{author}{\bibfnamefont{K.}~\bibnamefont{Furuya}}, \bibnamefont{and}
  \bibinfo{author}{\bibfnamefont{J.~H.} \bibnamefont{Lowenstein}},
  \bibinfo{journal}{Rev. Mod. Phys.} \textbf{\bibinfo{volume}{55}},
  \bibinfo{pages}{331} (\bibinfo{year}{1983}).

\bibitem[{\citenamefont{Furuya and Lowenstein}(1982)}]{FL82}
\bibinfo{author}{\bibfnamefont{K.}~\bibnamefont{Furuya}} \bibnamefont{and}
  \bibinfo{author}{\bibfnamefont{J.~H.} \bibnamefont{Lowenstein}},
  \bibinfo{journal}{Phys. Rev. B} \textbf{\bibinfo{volume}{25}},
  \bibinfo{pages}{5935} (\bibinfo{year}{1982}).

\bibitem[{\citenamefont{Parcollet and Georges}(1997)}]{PG97}
\bibinfo{author}{\bibfnamefont{O.}~\bibnamefont{Parcollet}} \bibnamefont{and}
  \bibinfo{author}{\bibfnamefont{A.}~\bibnamefont{Georges}},
  \bibinfo{journal}{Phys. Rev. Lett.} \textbf{\bibinfo{volume}{79}},
  \bibinfo{pages}{4665} (\bibinfo{year}{1997}).

\bibitem[{\citenamefont{Coleman and Paul}(2004)}]{CP04pre}
\bibinfo{author}{\bibfnamefont{P.}~\bibnamefont{Coleman}} \bibnamefont{and}
  \bibinfo{author}{\bibfnamefont{I.}~\bibnamefont{Paul}}
  (\bibinfo{year}{2004}), \bibinfo{note}{cond-mat/0404001}.

\bibitem[{\citenamefont{Mehta et~al.}(2004)\citenamefont{Mehta, Borda,
  Zar\'and, Andrei, and Coleman}}]{MBZAC04pre}
\bibinfo{author}{\bibfnamefont{P.}~\bibnamefont{Mehta}},
  \bibinfo{author}{\bibfnamefont{L.}~\bibnamefont{Borda}},
  \bibinfo{author}{\bibfnamefont{G.}~\bibnamefont{Zar\'and}},
  \bibinfo{author}{\bibfnamefont{N.}~\bibnamefont{Andrei}}, \bibnamefont{and}
  \bibinfo{author}{\bibfnamefont{P.}~\bibnamefont{Coleman}}
  (\bibinfo{year}{2004}), \bibinfo{note}{cond-mat/0404122}.

\bibitem[{\citenamefont{Pustilnik and Glazman}(2004)}]{PG04}
\bibinfo{author}{\bibfnamefont{M.}~\bibnamefont{Pustilnik}} \bibnamefont{and}
  \bibinfo{author}{\bibfnamefont{L.}~\bibnamefont{Glazman}},
  \bibinfo{journal}{J. Phys.: Condens. Matter} \textbf{\bibinfo{volume}{16}},
  \bibinfo{pages}{R513} (\bibinfo{year}{2004}).

\bibitem[{\citenamefont{Posazhennikova and Coleman}(2004)}]{PC04pre}
\bibinfo{author}{\bibfnamefont{A.}~\bibnamefont{Posazhennikova}}
  \bibnamefont{and} \bibinfo{author}{\bibfnamefont{P.}~\bibnamefont{Coleman}}
  (\bibinfo{year}{2004}), \bibinfo{note}{cond-mat/0410001}.

\bibitem[{\citenamefont{Cragg and Lloyd}(1979)}]{CL79}
\bibinfo{author}{\bibfnamefont{D.~M.} \bibnamefont{Cragg}} \bibnamefont{and}
  \bibinfo{author}{\bibfnamefont{P.}~\bibnamefont{Lloyd}}, \bibinfo{journal}{J.
  Phys. C} \textbf{\bibinfo{volume}{12}}, \bibinfo{pages}{L215}
  (\bibinfo{year}{1979}).

\bibitem[{\citenamefont{Wilson}(1975)}]{Wil75}
\bibinfo{author}{\bibfnamefont{K.}~\bibnamefont{Wilson}},
  \bibinfo{journal}{Rev. Mod. Phys.} \textbf{\bibinfo{volume}{47}},
  \bibinfo{pages}{773} (\bibinfo{year}{1975}).

\bibitem[{\citenamefont{Sakai et~al.}(1989)\citenamefont{Sakai, Shimizu, and
  Kasuya}}]{SSK89}
\bibinfo{author}{\bibfnamefont{O.}~\bibnamefont{Sakai}},
  \bibinfo{author}{\bibfnamefont{Y.}~\bibnamefont{Shimizu}}, \bibnamefont{and}
  \bibinfo{author}{\bibfnamefont{T.}~\bibnamefont{Kasuya}},
  \bibinfo{journal}{J. Phys. Soc. Japan} \textbf{\bibinfo{volume}{58}},
  \bibinfo{pages}{3666} (\bibinfo{year}{1989}).

\bibitem[{\citenamefont{Costi et~al.}(1994)\citenamefont{Costi, Hewson, and
  Zlatic}}]{CHZ94}
\bibinfo{author}{\bibfnamefont{T.~A.} \bibnamefont{Costi}},
  \bibinfo{author}{\bibfnamefont{A.~C.} \bibnamefont{Hewson}},
  \bibnamefont{and} \bibinfo{author}{\bibfnamefont{V.}~\bibnamefont{Zlatic}},
  \bibinfo{journal}{J. Phys.: Condens. Matter} \textbf{\bibinfo{volume}{6}},
  \bibinfo{pages}{2519} (\bibinfo{year}{1994}).

\bibitem[{\citenamefont{Hofstetter and Zar\'and}(2004)}]{HZ04}
\bibinfo{author}{\bibfnamefont{W.}~\bibnamefont{Hofstetter}} \bibnamefont{and}
  \bibinfo{author}{\bibfnamefont{G.}~\bibnamefont{Zar\'and}},
  \bibinfo{journal}{Phys. Rev. B} \textbf{\bibinfo{volume}{69}},
  \bibinfo{pages}{235301} (\bibinfo{year}{2004}).

\bibitem[{\citenamefont{Langreth}(1966)}]{Lan66}
\bibinfo{author}{\bibfnamefont{D.}~\bibnamefont{Langreth}},
  \bibinfo{journal}{Phys. Rev.} \textbf{\bibinfo{volume}{150}},
  \bibinfo{pages}{516} (\bibinfo{year}{1966}).

\bibitem[{\citenamefont{Schrieffer and Wolff}(1966)}]{SW66}
\bibinfo{author}{\bibfnamefont{J.~R.} \bibnamefont{Schrieffer}}
  \bibnamefont{and} \bibinfo{author}{\bibfnamefont{P.~A.} \bibnamefont{Wolff}},
  \bibinfo{journal}{Phys. Rev.} \textbf{\bibinfo{volume}{149}},
  \bibinfo{pages}{491} (\bibinfo{year}{1966}).

\bibitem[{\citenamefont{Sinjukow et~al.}(2002)\citenamefont{Sinjukow, Meyer,
  and Nolting}}]{SMN02}
\bibinfo{author}{\bibfnamefont{P.}~\bibnamefont{Sinjukow}},
  \bibinfo{author}{\bibfnamefont{D.}~\bibnamefont{Meyer}}, \bibnamefont{and}
  \bibinfo{author}{\bibfnamefont{W.}~\bibnamefont{Nolting}},
  \bibinfo{journal}{phys. stat. sol. (b)} \textbf{\bibinfo{volume}{233}},
  \bibinfo{pages}{536} (\bibinfo{year}{2002}).

\bibitem[{\citenamefont{Zar\'and et~al.}(2004)\citenamefont{Zar\'and, Borda,
  von Delft, and Andrei}}]{ZBDA04pre}
\bibinfo{author}{\bibfnamefont{G.}~\bibnamefont{Zar\'and}},
  \bibinfo{author}{\bibfnamefont{L.}~\bibnamefont{Borda}},
  \bibinfo{author}{\bibfnamefont{J.}~\bibnamefont{von Delft}},
  \bibnamefont{and} \bibinfo{author}{\bibfnamefont{N.}~\bibnamefont{Andrei}}
  (\bibinfo{year}{2004}), \bibinfo{note}{cond-mat/0403696}.

\bibitem[{\citenamefont{Hewson}(1993)}]{hewson}
\bibinfo{author}{\bibfnamefont{A.~C.} \bibnamefont{Hewson}},
  \emph{\bibinfo{title}{The Kondo Problem to Heavy Fermions}}
  (\bibinfo{publisher}{Cambridge University Press}, \bibinfo{year}{1993}).

\end{thebibliography}
\end{document}